\documentstyle[11pt,epsf]{article}
\begin{document}

\pagestyle{plain}
\pagenumbering{arabic}
\setcounter{section}{0}
\setcounter{subsection}{0}
\setcounter{subsubsection}{0}
\setcounter{equation}{0}
\setcounter{footnote}{0}
\setcounter{table}{0}
\setcounter{figure}{0}

% Useful macros
\newcommand{\etc}{{\em et al.}}
\newcommand{\etal}{{\em et al.}}
\newcommand{\cen}{\multicolumn{1}{c}}
\newcommand{\Sp}{S$_{\mbox{p}}$}
\newcommand{\gsim}{\ \mbox{\raisebox{-0.9ex}{$\stackrel{\textstyle>}{\sim}$}}\ }
\newcommand{\lsim}{\ \mbox{\raisebox{-0.9ex}{$\stackrel{\textstyle<}{\sim}$}}\ }
\newcommand{\eg}{{e.g.\/,}}
\newcommand{\ie}{{i.e.\/,}}
\newcommand{\quo}[1]{`#1'}
\newcommand{\kms}{\mbox{km~s$^{-1}$}}
\newcommand{\rv}{\mbox{$R_{vir}$}}
\newcommand{\mv}{\mbox{$M_{vir}$}}
\newcommand{\tv}{\mbox{$T_{vir}$}}
\newcommand{\uv}{\mbox{$u_{vir}$}}
\newcommand{\myref}[1]{\ref{#1}}
\newcommand{\eqn}[1]{Equation~\myref{#1}}
\newcommand{\tab}[1]{Table~\myref{#1}}
\newcommand{\fig}[1]{Figure~\myref{#1}}
\newcommand{\sect}[1]{Section~\myref{#1}}

% Observatory macros
\newcommand{\ROSAT}{{\em ROSAT}}
\newcommand{\Einstein}{{\em Einstein}}
\newcommand{\EXOSAT}{{\em EXOSAT}}
\newcommand{\Tenma}{{\em Tenma}}
\newcommand{\Ginga}{{\em Ginga}}

% Journal macros (for MNRAS)
\newcommand{\ASTRO}[1]{preprint (astro-ph/#1)}
\newcommand{\MNRAS}[2]{{ MNRAS\/}, {#1}, #2}
\newcommand{\ApJ}[2]{{ ApJ\/}, {#1}, #2}
\newcommand{\ApJS}[2]{{ ApJS\/}, {#1}, #2}
\newcommand{\ApJSub}{ ApJ, submitted}
\newcommand{\AnA}[2]{{ A\&A\/}, {#1}, #2}
\newcommand{\AJ}[2]{{ AJ\/}, {#1}, #2}
\newcommand{\ARAA}[2]{{ ARA\&A\/}, {#1}, #2}
\newcommand{\Nature}[2]{{Nat\/}, {#1}, #2}
\newcommand{\Science}[2]{{Sci\/}, {#1}, #2}

%\tableofcontents
%\newpage

\begin{center}
\section*{The Properties of the Hot Gas in Galaxy Groups and Clusters from 
           1-D Hydrodynamical Simulations -- I. Cosmological Infall
           Models \\[5mm]}
\end{center}
P.A. Knight and T.J. Ponman\footnote{To whom correspondence 
should be addressed.}\newline
School of Physics and Space Research, University of Birmingham, Edgbaston, 
Birmingham B15 2TT, UK \\[15mm]
%\hspace*{10cm}
%Draft: \today

\section*{ABSTRACT}
We report the results of 1-D hydrodynamical modelling of the evolution
of gas in galaxy clusters. We have incorporated many of the effects
missing from earlier 1-D treatments: improved modelling of the dark
matter and galaxy distributions, cosmologically realistic evolution of
the cluster potential, and the effects of a multiphase cooling flow. The
model utilises a fairly standard 1-D Lagrangian hydrodynamical code to
calculate the evolution of the intracluster gas. This is coupled to a
theoretical model for the growth of dark matter density perturbations.
The main advantages of this treatment over 3-D codes are (1) improved spatial
resolution within the cooling flow region, (2) much faster execution time,
allowing a fuller exploration of parameter space, and (3) the inclusion
of additional physics.

In the present paper, we explore the development of infall models -- in
which gas relaxes into a deepening potential well -- covering a wide
range of cluster mass scales. We find that such simple models reproduce
many of the global properties of observed clusters. Very strong cooling
flows develop in these 1-D cluster models. In practice, disruption by
major mergers probably reduces the cooling rate in most clusters.
The models overpredict the gas fraction in low mass systems, indicating
the need for additional physical processes, such as preheating or galaxy
winds, which become important on small mass scales.

\section{INTRODUCTION}
Rich clusters of galaxies have long been known to be strong sources of 
diffuse X-ray emission, caused by thermal bremsstrahlung from a hot
($T\sim10^8$K) intracluster plasma (the intracluster medium, or ICM). 
This gas is the dominant baryonic component in such systems, 
contributing 20--30\% of the gravitating material within an Abell radius.
More recently, it has become clear that such a diffuse component
is also present in poor clusters and galaxy groups.
A good understanding of the processes involved in the formation and 
evolution of this diffuse gas is therefore of key importance for any
complete theory of structure formation over a large range of scales. 

The earliest numerical models of the evolution of the intracluster gas
(Gull \& Northover 1975; Lea 1976; Takahara \etal\ 
1976) were all one-dimensional simulations, concerned primarily with the 
evolution of uniform density gas `clouds' relaxing into static 
potentials formed by the cluster galaxies and dark matter.
In this simple situation, infall of gas leads to a rise in central
gas density, and a shock propagates 
outwards from the centre of the cluster, heating the gas to  
temperatures similar to those observed today. After passage of the shock, 
the gas is approximately in hydrostatic equilibrium and further evolution 
is quasistatic. The subsequent cluster luminosity is found to be nearly constant, 
although Lea (1976) found large fluctuations in $L_X$, probably due to 
numerical problems. The resulting gas distributions are generally not 
well fitted by any polytropic distribution, although Gull \& Northover (1975),
who considered large radius infall into an initially empty potential, 
found that the cluster gas distributions were almost adiabatic. 

Perrenod (1978) considered more realistic infall models in which the cluster 
potential varies with time, using the results of an $N$-body simulation of the 
evolution of the galaxies in a Coma-like cluster (White 1976). In these 
simulations, the cluster galaxies collapse rapidly by violent relaxation, 
followed by a slower contraction due to two-body encounters. Unlike 
models with a static potential, the cluster X-ray luminosity was found to 
rise by an order of magnitude from $z=1$ to $z=0$ (for a flat Einstein-de 
Sitter cosmology), due primarily to the deepening of the potential well. 
However, this rate of evolution is considerably larger than 
that observed and may 
reflect overmerging in the galaxy distribution -- White's models assumed 
that all the cluster mass was bound to the galaxies, overestimating the 
strength of two-body encounters. 

Hirayama (1978) extended the purely primordial simulations and 
considered models including mass and energy injection from a static 
galaxy distribution. These simulations showed that the injected material 
dominated the ICM towards the centre, primarily because the gas distribution 
in the cluster was more extended than that assumed for the galaxies. For 
high gas injection temperatures ($T_{inj}\gsim3\times10^8$ K) a cluster 
wind formed.

More recently, much effort has been expended on the 
development of 3-D codes that are 
able to calculate the evolution of the gas and one or more collisionless 
components (e.g. dark matter, galaxies) from the primordial density 
fluctuation spectrum. This allows a fully self-consistent treatment in 
which the ICM evolves in a three-dimensional, time-varying potential. 
Since hierarchical clustering models will, for many 
interesting power spectra, lead to a high degree of substructure at all 
stages of the evolution, grid-based hydrodynamical techniques with their 
limited spatial resolution are unable to evolve the different levels of 
clustering hierarchy simultaneously. The gas flow is therefore generally 
calculated using particle-based techniques, for example smoothed particle 
hydrodynamics (SPH; Gingold \& Monaghan 1977), coupled with a
suitable $N$-body algorithm (\eg\ P$^3$M; Efstathiou \& Eastwood, 1981). 

Several authors have computed 3-D evolutionary models using such
techniques -- \eg\ Evrard (1988, 1990), Thomas \& Couchman (1992), Katz
\& White (1993), Metzler \& Evrard (1994), Navarro, Frenk \& White
(1996), hereafter NFW96. In most cases, cluster cores are found to
develop at the intersection of filaments and sheets, and cluster growth
proceeds through accretion of surrounding material as well as through
mergers with smaller, collapsed systems.

Although the general problem of non-linear evolution requires resort to
numerical techniques, it is possible to study the evolution of structure
in hierarchical models of the Universe by exploiting the approximate
self-similarity of the evolution expected under certain conditions:
namely those in which the background universe has closure density
($\Omega=1$), so the Universe is scale free, and in which the amplitude
of the initial density fluctuations follows a power-law, so these
perturbations are also scale free. In this situation, the clustering is
self-similar in the sense that statistical properties of the cluster
population must evolve according to simple scaling laws (Kaiser 1986).

In the idealised case of the spherically symmetric growth of an overdense
region within an environment in which the overdensity varies initially
as a power law function of radius, the evolution of {\it individual}
clusters develops in a self-similar fashion (Fillmore \& Goldreich 1984
(FG84); Bertschinger 1985). In this case, the profiles of a growing cluster
at different epochs are similar, and clusters of different
mass at a given epoch are simply scaled versions of one another.
Since it has been shown by Hoffman \& Shaham (1985) that the mean 
primordial overdensity distribution about a local peak does take a
power law form, one might expect such self-similar growth to give
a reasonable approximation to the mean growth of a cluster.

In practice, the similarity will be disturbed by departures from
spherical symmetry, and by interaction between neighbouring density
peaks. Nonetheless, simulations (e.g. Navarro, Frenk \& White 1995,
NFW96) indicate that when additional physics, such as radiative cooling
and injection of winds from galaxies, is not included, cluster growth
{\it is} approximately self-similar, with density profiles similar in
form to those predicted by infall models of cluster formation (FG84;
Bertschinger 1985). Also, clusters of different mass are very similar
when appropriately scaled -- though when a wide mass range is considered,
lower mass systems are found to be rather more centrally concentrated,
since they tend to form earlier. These results encourage the idea that
1-dimensional simulations should give a reasonable approximation to the
mean evolution of clusters.

Since previous 1-D treatments were too simplistic (none
combined a realistic evolving potential with gas injection and cooling), and 
the limits of present day computers mean that 3-D simulations
have poor resolution ($\gsim100$ kpc), 
numerical simulations have thus far been unable to shed much light on a 
number of unanswered questions: \eg\ Do the scaling laws break down when 
cooling is included? Were cooling flows more massive in the past? How 
does the mass deposition rate vary with radius and system size?
How does the gas mass fraction vary with radius? How does the cluster 
luminosity vary with time? 

The purpose of the present study is to address some of these questions,
using a 1-D code, {\em EVOL}, that incorporates many of the effects
missing from the older 1-D treatments: improved modelling of the dark
matter and galaxy distributions, more realistic gas injection mechanisms,
the effects of convection and multiphase cooling, as well as a simple
investigation of the effects of mergers.

The model utilises a fairly standard 1-D Lagrangian hydrodynamical code
to calculate the evolution of the intracluster gas. This code is coupled
to a theoretical model for the growth of dark matter density
perturbations, which gives a cluster potential that deepens in a
cosmologically realistic way with time. The main advantages of this
treatment over 3-D codes are (1) improved spatial resolution within the
cooling flow region, (2) fast execution time, allowing a fuller
exploration of parameter space, and (3) the inclusion of additional
physics. The main disadvantage is that we are unable to accurately model
the effects of asymmetric processes, such as cluster mergers. This
limitation, however, does not invalidate our results since, although the
evolution of the dark matter in clusters is complex, with many clusters
showing evidence of substructure, the dark matter distribution in
``quiescent'' clusters can be reasonably well approximated by a
spherically symmetric model. For example,
recent simulations by Tormen, Bouchet \& White (1996)
show that dark matter halos spend about two thirds of their evolution in
a relaxed state, during which the density profiles of the halos are well
approximated by the NFW96 analytical model. Having said this, periodic
major mergers may well have a substantial impact on the structure within
the central cooling flow region, possibly disrupting the flow. We
investigate this effect below, using a simple model for the change in
entropy of the gas within the cooling flow following a merger.

Our aim is to determine the role of, and place
constraints on, the heating and cooling processes that are important for
the evolution of the ICM. The current paper deals with models in which
the intracluster gas is unaffected by injection of energy and heavy
elements from cluster galaxies, in order to test the extent to which the
observed properties of clusters and groups can be explained by models
with gas dynamics and radiative cooling, without recourse to additional
physics. Models with gas injection from the galaxies are considered in a
second paper (Paper II -- Ponman and Knight, in preparation).

The layout of the paper is as follows: the equations on which the
code is based, and the numerical methods employed, are outlined in
\sect{method}, and some test runs used to verify the accuracy of the code
are described in \sect{tests}. A large number of simulations have been
run, in order to investigate the mass dependence of the evolution,
and to check the effect of varying a number of key physical parameters.
This grid of models is described in \sect{grid}, and the evolution of the
hot gas which results is discussed in broad terms in \sect{soltns}. 
Some more detailed aspects of the properties of these simulated clusters
are then investigated: cooling flows in \sect{cf}, the distribution of hot
gas in \sect{fgas}, and the evolution of the X-ray luminosity in
\sect{lum}. The simulations presented here assume an $\Omega=1$ Universe,
however in \sect{omega} we examine briefly the impact of a lower value of
$\Omega$ on the X-ray evolution, and conclude that it is not large.
Finally, in \sect{conc}, we summarise our conclusions.

\section{NUMERICAL METHOD\label{method}}

\subsection{Basic Equations\label{describe}}
Assuming spherical geometry, the evolution of the ICM in galaxy clusters is 
described by the standard time-dependent equations of gas dynamics 
(Mathews \& Baker 1971), with additional terms to remove gas from the flow when 
the cooling time is sufficiently short in order to simulate the features of 
an inhomogeneous cooling flow. Under such conditions, the 
equations of mass, momentum, and energy conservation that govern the 
evolution of the ICM are
\begin{equation}
  \frac{\partial\rho}{\partial t}+\frac{1}{r^2}
  \frac{\partial(r^2\rho u)}{\partial r}=-\omega\rho,\label{mass}
\end{equation}
\begin{equation}
  \frac{Du}{Dt} = -\frac{1}{\rho}\frac{\partial(p+q)}{\partial r}+g,\label{motion}
\end{equation}
\begin{equation}
  \frac{D\epsilon}{Dt}-\frac{p+q}{\rho^2}\frac{D\rho}{Dt}
  =-\frac{n_e n_H}{\rho}\Lambda,\label{energy}
\end{equation}
where $D/Dt$ is the Lagrangian derivative, $u$ 
the radial gas velocity (positive for outflow), $\rho$ the 
ICM mass density, $n_e$ the electron number density, $n_H$ the hydrogen
number density, and $\omega$ accounts for gas lost by cooling (see section
2.2 below). For a gas with ratio of specific heats $\gamma=5/3$, the specific 
energy $\epsilon$ and the gas pressure $p$ are equal to $3kT/(2\mu m_p)$ and 
$\rho kT/(\mu m_p)$ respectively, where $T$ is the gas temperature, $k$ is 
the Boltzmann constant and $m_p$ is the proton mass. Throughout this paper 
we take $\mu=0.6$ and $\mu_e=1.15$, appropriate for a fully-ionized plasma 
composed of 90\% hydrogen and 10\% helium by number. 

The term $q$ in Equations~\ref{motion} and~\ref{energy} is the artificial 
viscosity (Richtmyer \& Morton 1967),
\begin{equation}
   q=\rho\left(a\min\left\{\frac{\partial u}{\partial r},0\right\}\right)^2,
   \label{q}
\end{equation}
which is effective at smoothing the boundaries of shock waves. The parameter 
$a$ determines the thickness of the shock front. Setting $a=2$, the value that 
is adopted throughout this paper, ensures accurate modelling of the shock by
spreading the jump in parameter values over several Lagrangian shells.

\subsection{Radiative Cooling and Mass Deposition}
The term on the right hand side of the energy equation accounts for the effect 
of radiative cooling. $\Lambda$ is evaluated using bilinear interpolation from 
tabulated versions of the cooling function of Raymond, Cox, \& Smith 
(1976) for primordial and solar abundances, 
\begin{equation}
\Lambda\left\{T,Z\right\}=(1-Z)\Lambda\left\{T,0\right\}+Z\Lambda\left\{T,1\right\},
\end{equation}
where $Z$ is the metallicity of the gas. Compton cooling is only
important at high redshift ($z\gsim10$), and is ignored 
in the simulations presented here. At such early epochs additional energy
input from early star formation is likely to be effective at preheating the
cluster gas. The effects of this will be explored in Paper II.

Observed X-ray surface brightness profiles in cooling flows imply that
gas cools and is deposited throughout the flow (Fabian, Nulsen \&
Canizares 1991). The treatment of mass deposition used in {\em EVOL} is
similar to that of David, Forman \& Jones (1990). Gas is assumed to be
thermally unstable to entropy perturbations when $(2-d\ln\Lambda/d\ln
T)>0$ (David, Forman \& Jones 1990). When this condition is satisfied,
and the integrated isobaric cooling time, $t_{cool}$ is less than 
$\alpha_{md}t$, gas is removed from the ICM at
the rate $\xi\rho/t_{rad}$, where $\rho$ is the gas density, $t_{rad}$ 
is the local instantaneous cooling time, and $\alpha_{md}$ and $\xi$ are 
model parameters. Theoretical considerations
and observations of the cooling flows in rich clusters suggest $\xi\sim1$
(Sarazin 1990). This accounts for the last term in \eqn{mass},
\begin{equation}
  \omega=\left\{ \begin{array}{ll}
  \xi/t_{rad}, & \mbox{if $(2-d\ln\Lambda/d\ln T) > 0$ and $t_{cool}<\alpha_{md}t$} \\      
  0, & \mbox{otherwise} \end{array} \right.
\end{equation}

\subsection{The Cluster Potential}
The gravitational acceleration ($g$ in \eqn{motion}) is
given by,
\begin{equation}
  g=-\frac{G}{r^2}
  \left[M_{dark}(r)+M_{ICM}(r)+M_{dep}(r)\right],
\end{equation}
where $M_{dark}(r)$, $M_{ICM}(r)$ and $M_{dep}(r)$ are
the total dark matter, ICM and `cooled' gas mass within radius $r$
respectively. The cooled gas mass, $M_{dep}$, is primarily used for
book-keeping, keeping track of material that is lost from the ICM due to
mass deposition, although in the simulations it was found to be dynamically
important close to the cluster centre at later times. The dark matter
distribution otherwise dominates the dynamical evolution of the ICM. The
galaxy distribution is only important in models with significant
feedback from galaxy winds and ram-pressure stripping and is not considered
here. (The galaxies are effectively incorporated into the dark matter
component.)

The evolution of the dark matter in clusters is complex, with many
clusters showing evidence for recent mergers. However, the dark matter
distribution in quiescent clusters is reasonably well approximated by a
spherically symmetric model. NFW96 found that the radial dark matter
density profiles of their simulated clusters were approximately
self-similar, with profiles at large radii similar in form to that
predicted by analytic infall models (FG84; Bertschinger 1985). In these
models, the physical mass distribution of a cluster growing by the 1-D
accretion of surrounding material at time $t$ may be calculated by
scaling a canonical function to the characteristic size ($\rv(t)$) of the
system. We assume that such models give a good approximation to the shape
and evolution of the dark matter distribution of real clusters, and adopt
the canonical profile given by the similarity solution of FG84
corresponding to $n=-1$ (see Hoffman \& Shaham 1985), which matches the
slope of the CDM power spectrum on cluster scales (Kaiser 1986). This
model tends towards the critical background density $\rho_c$ at large
radii. Cusps in the density arising from the turnaround of particle 
shells are present in the similarity solution. In practice, such features
would be washed out by departures from perfect spherical symmetry -- we
therefore eliminate them by fitting a power law profile to the formal
similarity solution.

Following the $n=-1$ similarity solution,
the full dark matter profile at $z=0$ is determined by a single parameter,
which may be taken to be the total mass inside the virial radius \rv\
(defined throughout this work as the radius within
which the mean density of the system is 200$\rho_c$). The profile at
any earlier epoch is then simply a self-similar scaling of this;
$\rho/\rho_c$ being a constant function of $r/\rv$.

There is evidence from N-body simulations that the self-similar model
breaks down at small radii, where some flattening of the density
distribution is observed. Following the results of NFW96, we therefore
flatten the canonical density profile to a logarithmic slope of -1 within
a radius of $\eta_{\rho}\rv$, where $\eta_{\rho}$ is a selectable model
parameter. The gas is allowed to evolve within this developing dark
matter distribution, and the total gravitational potential is computed
from the combined contributions of dark matter, free gas, and gas
deposited by cooling.

\begin{figure}
\begin{center}
\leavevmode
\epsfxsize=8cm
\epsfbox[0 0 464 464]{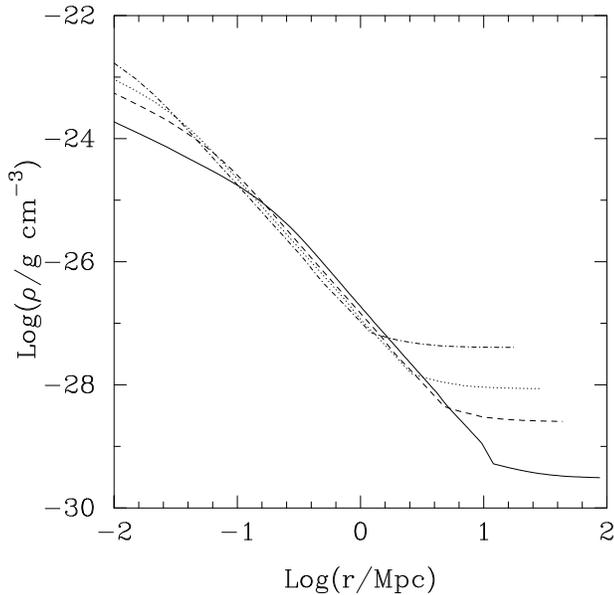}
\caption{Evolution of the dark matter density in a model with
$\mv=1.6\times10^{15}$ $M_{\odot}$. The profile evolves in a self-similar
way through redshifts 4 (dot-dash), 2 (dotted), 1 (dashed) and 0 (solid).
\label{dmevol}}
\end{center}
\end{figure}

The evolution of the dark matter distribution in physical units,
for the case of a cluster with virial mass $\mv=1.6\times10^{15}$
$M_{\odot}$ at $z=0$, is shown
in \fig{dmevol}. In scaled units, $\rho/\rho_c$ versus $r/\rv$,
these curves would all fall on top of one another.

Our approach contrasts with 1-D methods which are based on numerical
integration of the equations of motion governing the evolution of
collisionless dark matter shells (Thoul \& Weinberg 1995). While such
methods are more self-consistent than our approach (in the sense that the
effect of the gas distribution can be included in the calculation of the
dark matter dynamics) there is a high price to pay in terms of the
execution speed of the code, since a large number of dark matter shells
is required to avoid spurious shocks arising in the gas due to
fluctuations in the mass distribution. Our approach has the advantage of
avoiding this computational overhead whilst yielding a good approximation
to the expected evolution -- at large radii the evolution calculated by
numerical integration will be similar to that given by the analytic
solution since the gas and dark matter are distributed similarly. Only at
relatively small radii, where the gas and dark matter distributions can be
significantly different, are the two methods likely to differ, but it is
precisely at this point that 3-D simulations suggest that 1-D treatments
break down, presumably due to the effect of merging, and the dark matter
density profile flattens. We are able to incorporate this effect into our
model heuristically, by imposing a core on our dark matter distribution.

Equations~\ref{mass}--\ref{energy} are solved using a standard,
first-order, explicit finite-difference Lagrangian scheme (Richtmyer \&
Morton 1967), with additional terms to account for gravity, cooling,
and mass deposition. A useful description of the use of a Lagrangian code
in relation to cluster cooling flows is given by Thomas (1988). Full details
of our method are given in Knight (1996).

\section{TESTS OF THE CODE\label{tests}}
The code was subjected to a variety of tests, including comparison with
analytically soluble problems, previous hydrodynamical simulations, and
with the results of another 1-D Lagrangian code, kindly supplied by Peter
Thomas. Full details of these checks are given in
Knight (1996). Here we show the results of two of the most important
tests -- comparisons with the analytical evolution derived for a simple
1-D problem by Bertschinger, and with the results of 3-D simulations
which incorporate the effects of hierarchical merging.
%In addition, the numerical
%method ensures that there is no net mass loss or gain within the simulated 
%volume, and it has been confirmed that energy is conserved to better than 1\%.
   
\subsection{Bertschinger Infall Model}
One of the most powerful tests for a hydrodynamical code is the simulation
of a self-similar system whose behaviour is known analytically.
Such a comparison is useful since any departures of the simulation from 
the self-similar behaviour will be attributable to a combination of 
numerical effects and possibly to differences between the model and the 
simulated system (\eg\ different boundary conditions).  

\begin{figure}
\begin{center}
\leavevmode
\epsfxsize=14cm
\epsfbox[0 0 464 244]{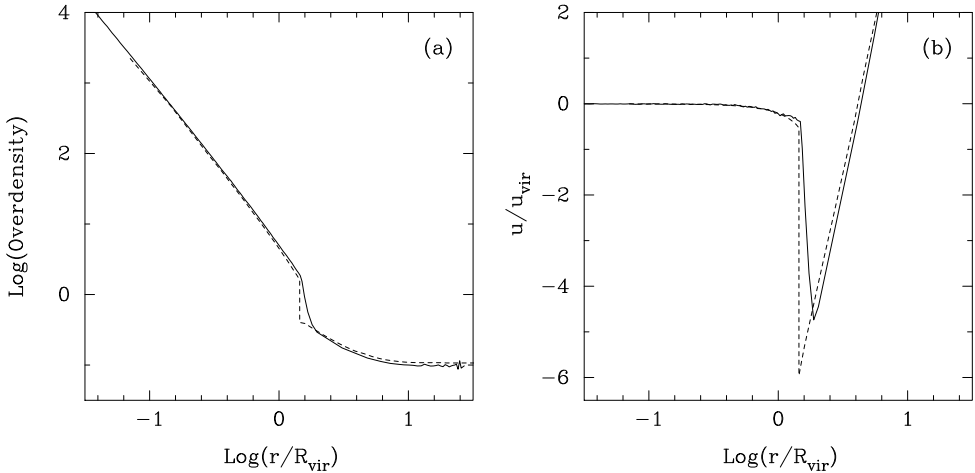}
\caption{Gas density and velocity profiles from {\em EVOL} (solid lines)
  compared with those predicted by self-similar theory (dashed lines;
  Bertschinger 1985). \rv\ is the virial radius and $\uv\equiv\rv/t$.
  The gas (baryon) fraction was taken to be 0.1.\label{bert}}
\end{center}
\end{figure}

Since the evolution of the dark matter profile is given by the infall
solution of FG84, in the absence of cooling and injection terms the
evolution of the gas in the case of a point perturbation ($n=0$) will
follow that derived by Bertschinger (1985). \fig{bert} shows the scaled
gas density and velocity profiles for an {\em EVOL} model with $n=0$
compared with that predicted analytically.
Clearly, there is excellent agreement between the profiles. The only
significant differences occur in the shock region, which is
spread out over several shells in the {\em EVOL} solutions.

\subsection{Comparison with a 3-D Code}
\begin{figure}
\begin{center}
\leavevmode
\epsfxsize=14cm
\epsfbox[0 0 464 244]{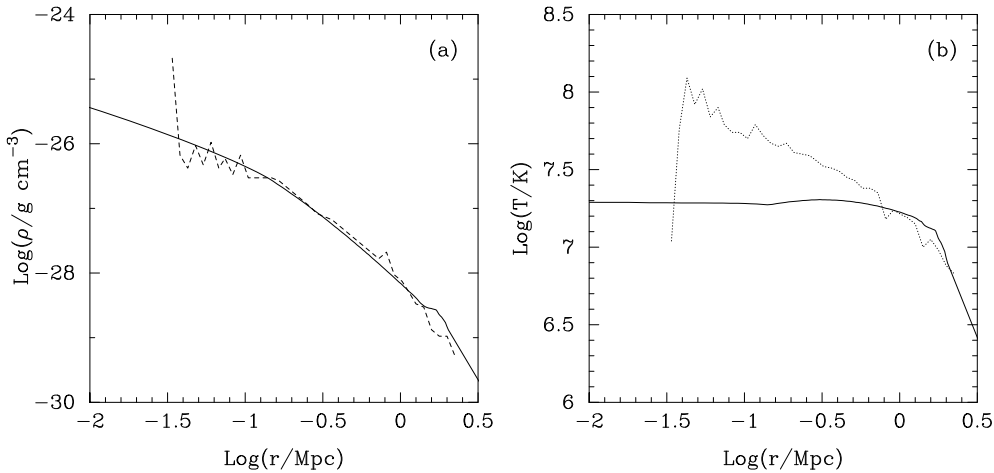}
\caption{Gas density and temperature profiles calculated by {\em EVOL} compared with 
  those calculated by a 3-D code (dotted line; Katz \& White 1993).\label{kw}}
\end{center}
\end{figure}

We have also performed some tests to investigate the differences between
{\em EVOL} calculations and those of a 3-D hydro code (Katz \& White
1993). For these simulations of an $\Omega=1$ universe with a CDM initial
power spectrum (modelled as an $n=-1$ self-similar hierarchy in the case
of {\em EVOL}), 9\% of the mass was taken to be in the form of gas, the
rest in dark matter. Radiative cooling was included.  In the 3-D
calculation, a cluster was found to form at the intersection of
filaments, forming an object with virial radius $R_{vir}=880$ kpc and
mass $M_{vir}=1.83\times10^{14}$ $M_{\odot}$ by $z=0.13$ when the
simulation was halted. The gas density and temperature of this structure
is compared to that calculated by {\em EVOL} in \fig{kw} -- the initial
conditions were taken to be as close to those of the 3-D code as
possible, and the normalisation of the initial perturbation was
determined from the virial mass of the final object in the 3-D
simulation.

There is excellent agreement in the shape of the gas density profile at
essentially all radii, apart from the sudden increase in the density at
the centre in the 3-D case due to the central `cooling catastrophe'
(which is avoided in {\em EVOL} by the distributed gas deposition).
The temperature profiles agree quite well for $r\simeq1$--3 Mpc, but that
from the 3-D simulation is much more centrally peaked than that from {\em
EVOL}. This is due to the rapid deepening of the potential well as gas
near the centre clumps under its own self-gravity, becomes even denser
and cools even more rapidly, leading to the formation of a much deeper
cluster potential in the 3-D case.  However, Katz and White comment that
the effect would be reduced if star formation was included (gas can cool
and stars cannot, reducing the rate at which the potential well can
deepen), which is effectively what is modelled by the multiphase
treatment employed by {\em EVOL}. Such central temperature gradients are 
typically absent in 3-D treatments in which cooling is neglected
(\eg\, Evrard 1990; NFW96),
most of which yield simulated clusters in which the central
regions are essentially isothermal (in general agreement with
observation, as discussed in \sect{soltns}).
Given the high sensitivity of the
results from such treatments to details of the model, we consider the
agreement between {\em EVOL} and 3-D codes to be satisfactory.

\section{THE MODEL GRID\label{grid}}
One of the main advantages of a 1-D treatment is the fast execution time,
enabling a much fuller exploration of parameter space than would be
possible with a 3-D code. This is very useful since the possible
parameter space for models of the intracluster medium is large, and many
of the key parameters that can influence the evolution of the ICM are
uncertain. In addition, we are especially interested in the way that the
properties of the ICM scale with system size. We have therefore
constructed a multidimensional grid that encompasses these parameters, as
well as a large range in system size, and calculated the evolution of
models at a large number ($\sim150$) of points in the parameter space.

Each model starts with 200 radial shells at $z=100$, early enough that
the assumed initial gas profiles have negligible effect on the profiles
for $z\lsim10$. The gas density and velocity were set to
$\rho(r)=f_{gas}\rho_{dark}(r)$, and $u(r)=H_ir$ (\ie\ comoving with the
Hubble flow) respectively, where $\rho_{dark}(r)$ is the initial dark
matter density profile (following FG84)
and $H_i$ is the Hubble constant at the initial
epoch. In practice, if a dark matter concentration has formed, then
the initial gas velocity will actually be somewhat perturbed from
the Hubble expansion. However, by starting at $z=100$, we ensure that
this has a negligible effect on the evolution.
The initial gas temperature was taken to be low ($T(r)=10^5$ K),
ensuring that the initial entropy of the gas is much lower than that
typically found in the cores of groups and rich clusters. A large initial
entropy for the gas would effectively introduce a scale size into the
evolution, leading to the formation of a gas density core. The
iron abundance of the gas was set to $Z(r)=0.4$, consistent with observed
mean abundances. In effect, we assume in the present set of models that
metals are injected into the ICM at some very early stage in cluster
evolution  --  for example, by Population III stars (e.g. Carr 1994;
Ostriker \& Gnedin 1996).

The initial dark matter and gas mass fractions for most of the runs
were set to $f_{dark}=0.8$ and $f_{gas}=0.2$ respectively. This represents
a baryon fraction of $f_b\simeq0.2$, higher than 
that generally used for cluster simulations ($f_b\simeq0.1$), and also
conflicts (if $\Omega=1$) with the current restrictive constraints on the 
universal baryon density parameter from primordial nucleosynthesis calculations,
$\Omega_b=0.05\pm0.01$ $h_{50}^{-2}$ (Walker \etal\ 1991). However, these
constraints are also in severe conflict with the observed baryon fractions in 
clusters of galaxies, 10--20\% at 1 Mpc (White \& Fabian 1995), rising to
$\simeq$30\% at the virial radius (David, Jones \& Forman 1995). Various possible
explanations for this difference have been proposed (White \etal\ 1993), 
some of which appeal to $\Omega_0\simeq0.3$, while others suggest that, 
within an $\Omega=1$ framework, baryons were preferentially concentrated 
within clusters at an early epoch, as a result of an unknown, non-gravitational 
mechanism. We take the view that, until this issue is resolved, it is
preferable to set the baryon fraction in the protoclusters to $\approx 0.2$, rather 
than produce models in which present day clusters are considerably underdense
in baryons compared to observation. 
However, some low baryon density models with $f_{dark}=0.9$ and $f_{gas}=0.1$
have also been investigated in order to bracket the observed baryon fraction 
range in clusters and groups. In addition, we have constructed some
models to explore the effects of a lower density Universe ($\Omega_0=0.3$)
on the evolution.

By varying the mass enclosed within the virial radius (\mv) at $z=0$
between $7.8\times10^{12}$ $M_{\odot}$ and $1.6\times10^{15}$
$M_{\odot}$, the grid covers a range of systems from groups with
$\tv\lsim1$ keV to rich clusters. N-body simulations suggest that the
radial infall calculation of the dark matter evolution is unphysical at
small radii. NFW96 find that the density
profile flattens to a logarithmic slope of approximately -1 near the centre. Our
mass profiles were therefore smoothly flattened to $\rho(r)\propto
r^{-1}$ within a ``core radius'' of $0.1\rv$, in accordance with their result. 
The results presented
here are relatively insensitive to the value of $\eta_{\rho}$, since cooled
gas generally becomes the dominant gravitational component at small
radii.

The mass deposition parameter was set to $\xi=1.7$ for shells where 
$t_{cool}<\alpha_{md}t$, and $\xi=0$ otherwise. Within the context 
of the mass deposition model, values of $\xi\sim1$ have both some theoretical 
and observational justification: if $\xi$ is too large the gradient 
of the specific entropy changes sign and the region becomes convectively 
unstable, effectively reducing $\xi$ back to $\sim$1 (Fabian, Nulsen \& 
Canizares 1991). The value $\xi=1.7$ was chosen since it was found to lead to
the formation of simulated deposition profiles of the form $\dot{M}(r)\propto r$,
similar to the form inferred from deprojection analyses of cluster cooling flows 
(Fabian, Nulsen \& Canizares 1991). However, some simulations with $\xi=1$ and
$\xi=2.5$ have also been performed.

The mass deposition cutoff at $t_{cool}>\alpha_{md}t$ is somewhat
arbitrary, although if deposition were to occur with $\xi\sim1$ out to
large radii, the resulting cluster spectra conflict with
observations -- for example, {\em Ginga} observations of the Perseus cluster
appear to rule out all but a modestly multiphase ICM (Allen \etal\ 1992).
$\alpha_{md}=1$ was used for most of the runs, although the effect of
confining mass deposition to the region $t_{cool}<0.2t$ has also been
explored. Mass deposition reduces the mass of shells within the cooling
flow region, helping to improve the resolution of the runs -- shells are
dropped when their mass falls below $10^6$ $M_{\odot}$, and any remaining
mass is added to a static Eulerian grid which is included in the
calculation of the cluster potential.
\begin{table}
\begin{center}
\begin{tabular}{llll} \hline \hline \\
\cen{Parameter} & \cen{Name} & \cen{Units} & \cen{Values} \\ \\ \hline \\
Density parameter     & $\Omega_0$ & --- & 0.3, 1 \\[2mm]
Gas mass fraction & $f_{gas}$ & --- & 0.1, 0.2 \\[2mm]
Final virial mass & $\mv$   & $M_{\odot}$ & $7.8\times10^{12}$, $2.0\times10^{13}$, 
   $3.9\times10^{13}$, $7.8\times10^{13}$, \\
   & & & $2.0\times10^{14}$, $3.9\times10^{14}$, $7.8\times10^{14}$, 
   $1.6\times10^{15}$ \\[2mm]
Mass deposition rate & $\xi$ & --- & 1, 1.7, 2.5 \\[2mm]
Mass deposition threshold & $\alpha_{md}$ & --- & 0.2, 1, $\infty$ \\ \\ \hline
\end{tabular}
\caption{Summary of the range of model parameters.\label{modpar}}
\end{center}
\end{table}

\section{MODEL EVOLUTION\label{soltns}}
The parameter space for the cluster models has many dimensions. It is therefore
convenient to adopt standard values for the model parameters and then test the
effect of alternative values. These standard values are taken to be
(1) $\Omega_0=1$, (2) gas fraction $f_{gas}=0.2$, (3) mass deposition rate $\xi=1.7$,
and (4) mass deposition threshold $\alpha_{md}=1$.  In the following discussion  
attention is focussed on parameters that differ from these standard values.
Parameters not explicitly mentioned are assumed to have the `standard' value.

\begin{figure}
\begin{center}
\leavevmode
\epsfxsize=14cm
\epsfbox[0 0 543 790]{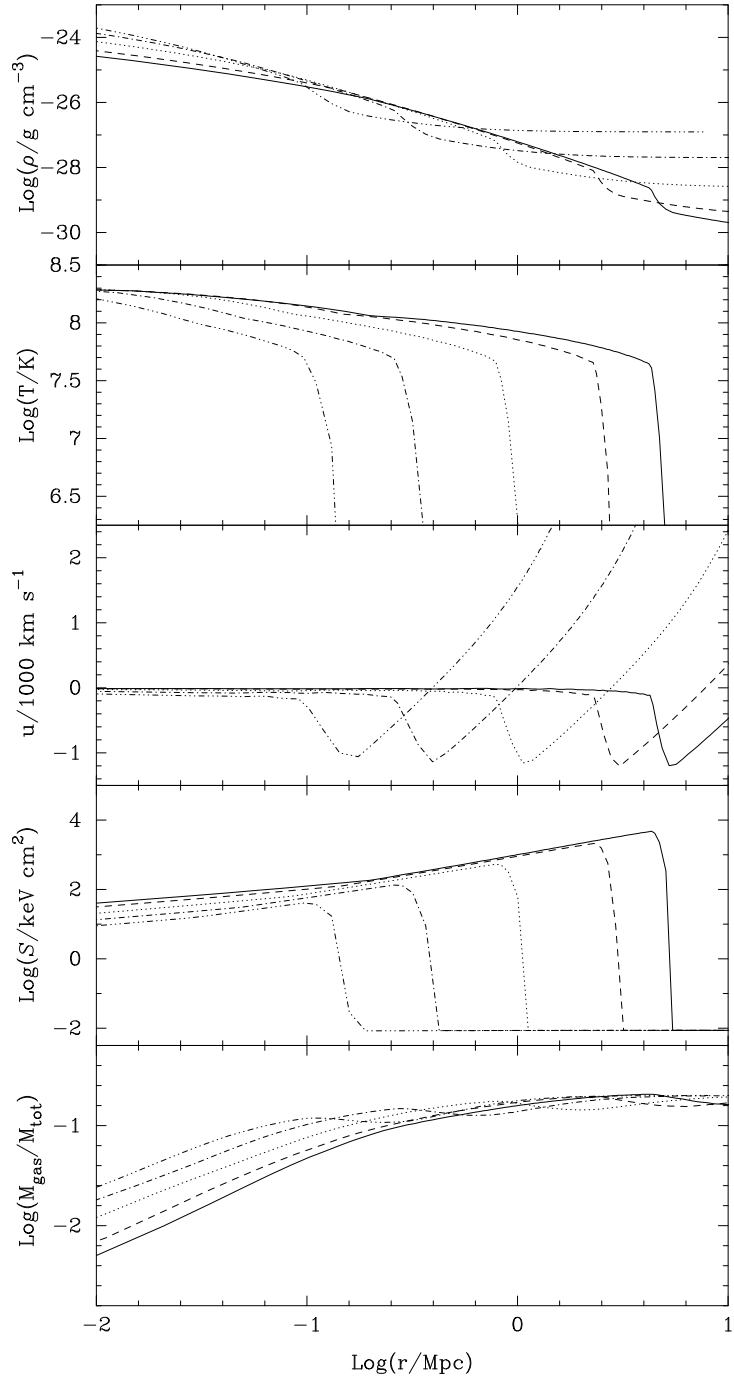}
\caption{Evolution of the gas density, temperature, velocity, 
  entropy, and enclosed gas mass fraction of a standard model of
  mass $\mv=1.6\times10^{15}$ $M_{\odot}$. 
  Redshifts shown are 0, 0.5, 2, 5 and 10 (solid, dashed, dotted, 
  dash-dot-dash, and  dash-dot-dot-dot-dash lines respectively).
  \label{inf1}} 
\end{center}
\end{figure}

\begin{figure}
\begin{center}
\leavevmode
\epsfxsize=14cm
\epsfbox[0 0 543 790]{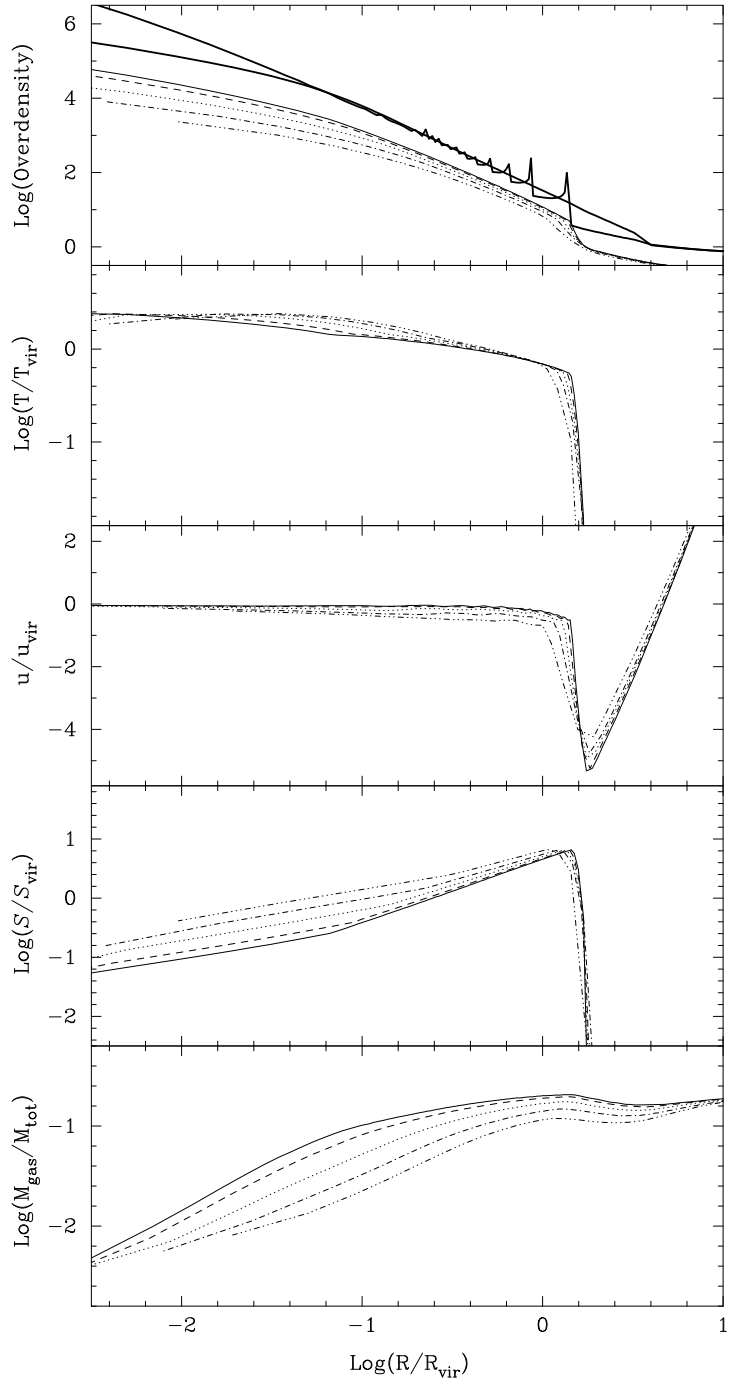}
\caption{Evolution of the scaled gas density, temperature, velocity, entropy,
  and enclosed gas mass fraction for a standard model of mass 
  $\mv=1.6\times10^{15}$ $M_{\odot}$.  Redshifts shown are 0, 0.5, 2, 5 and 10 
  (solid, dashed, dotted, dash-dot-dash, and dash-dot-dot-dot-dash lines 
  respectively). The thick solid lines in the top panel show the 
  ``unsmoothed'' dark matter density profile (with cusps), and the profile 
  represented by a power law  with an $r^{-1}$ core applied
  (see section 2.3).\label{inf2}}
\end{center}
\end{figure}

The evolution of a standard model of mass $\mv=1.6\times10^{15}$ $M_{\odot}$
in physical and ``scaled'' units is shown in Figures~\ref{inf1} 
and~\ref{inf2} respectively. The scaled plots 
were obtained by normalising to the characteristic temperature, velocity and 
entropy: $\tv=\mu m_p G\mv/(2 k \rv)$, $u_{vir}=\rv/t$, and
$S_{vir}=\tv/\rho_{vir}^{\gamma-1}$, where $\rv$ is the radius 
within which the mean density, $\rho_{vir}$, is 200 times the 
background density at time $t$. The characteristic temperature, $\tv$,
is $8.0\times10^7$ K, $3.2\times10^7$ K, $1.1\times10^7$ K, and $4.4\times10^6$ K
for models of mass $\mv=1.6\times10^{15}$ $M_{\odot}$, $3.9\times10^{14}$ $M_{\odot}$,
$7.8\times10^{13}$ $M_{\odot}$, and $2.0\times10^{13}$ $M_{\odot}$ respectively.
$\tv$ is independent of redshift for $n=-1$.

The evolution at large radius is similar to that of the self-similar solution 
(\eg\, Bertschinger 1985). A particle within the overdense region 
eventually separates from the Hubble 
expansion, reaching turnaround, $u=0$, before collapsing back 
towards the centre. Before reaching $r=0$, the fluid particle 
passes through a shock where the velocity changes discontinuously, 
after which it settles to a constant fraction of its radius at turnaround.
In contrast, three dimensional simulations suggest that the kinetic energy of
fluid particles is not completely thermalised after passing through the
shock. However, the kinetic pressure is only $\sim$10\% of the thermal pressure
(Evrard 1990) and does not affect the results presented here
greatly. The position of the shock is marked by a sharp increase in 
gas entropy. Within the shock radius, the entropy rises continuously with
radius, $S\propto r^{1.3}$, and convection is suppressed.

The density profile, while following the dark matter profile quite well
at large radius is strongly affected by cooling and mass deposition
nearer the centre. Gas dropped near the cluster centre is replaced by gas
of higher entropy from further out, lowering the central gas density.
This, coupled with the build-up of a large mass of cooled material
produces a strong gradient in the enclosed gas mass fraction. At $z=0$,
the gas mass fraction within 0.5 Mpc, 1 Mpc and 3 Mpc is 0.12, 0.16 and
0.20 respectively. The logarithmic slope of the density profile is found
to flatten, from approximately -2 to -1, at a radius of $\sim$100 kpc,
somewhat less than typically observed core radii in rich clusters, 
$r_c\simeq 250$ kpc --
although many of the systems with the largest core radii may be recent
merger remnants. As expected, the effects of cooling are most dramatic at
early epochs, when the gas density is higher and the effect of mass
deposition consequently more widespread, leading to a reduction in the
gas mass fraction throughout the cluster. At later times, cooling is only
important near the cluster centre.

Figure~\ref{inf1} shows that the temperature generally declines
continuously with increasing radius, even within the cooling flow region.
This is not unexpected since the ``hot phase'' of such a flow is
effectively being modelled (the cooling gas is dropped from the ICM at
each timestep), and the gas will be adiabatically compressed as it flows
inward. Recent studies of multiphase cooling flows within mass profiles
with small cores confirm that the temperature of the hot phase will
remain approximately constant within the cooling region (Waxman \&
Miralda-Escud\'{e} 1995), while if the gravitating matter distribution is
very sharply peaked, as found here, the temperature is expected to rise
(White \& Sarazin 1987). Beyond this region, the temperature gradient
steepens with radius until the sudden drop at the shock. 
Similar temperature profiles have been found in a number of three
dimensional hydrodynamical simulations (e.g. NFW96).

Observations of the temperature structure of clusters have been mostly
restricted to the nearest and brightest systems until recently.
Observations with coded mask and collimated instruments indicate that
the Coma cluster (Hughes, Gorenstein \& Fabricant 1988; Watt \etal\
1992) has an approximately isothermal core within
$\sim$1 Mpc of the cluster centre with a steep temperature decline at 
large radius, whilst the Perseus cluster (Eyles \etal\ 1991) shows a modest
temperature decline outside the cooling region, with the
emission well modelled by a power-law temperature profile, $T(r)\propto
r^{-0.30\pm0.08}$. Recent results from ROSAT and ASCA (e.g. Allen, Fabian
\& Kneib 1996; Markevitch \& Vikhlinin 1996; Loewenstein \& Mushotzky
1996) suggest that most reasonably relaxed clusters have fairly flat
temperature distributions within the central $\sim$1~Mpc (apart from the
common presence of a central cooling flow), although steep gradients
at larger radii appear to be present in at least some high temperature
clusters (Markevitch 1996).

\begin{figure}
\begin{center}
\leavevmode
\epsfxsize=14cm
\epsfbox[0 0 543 790]{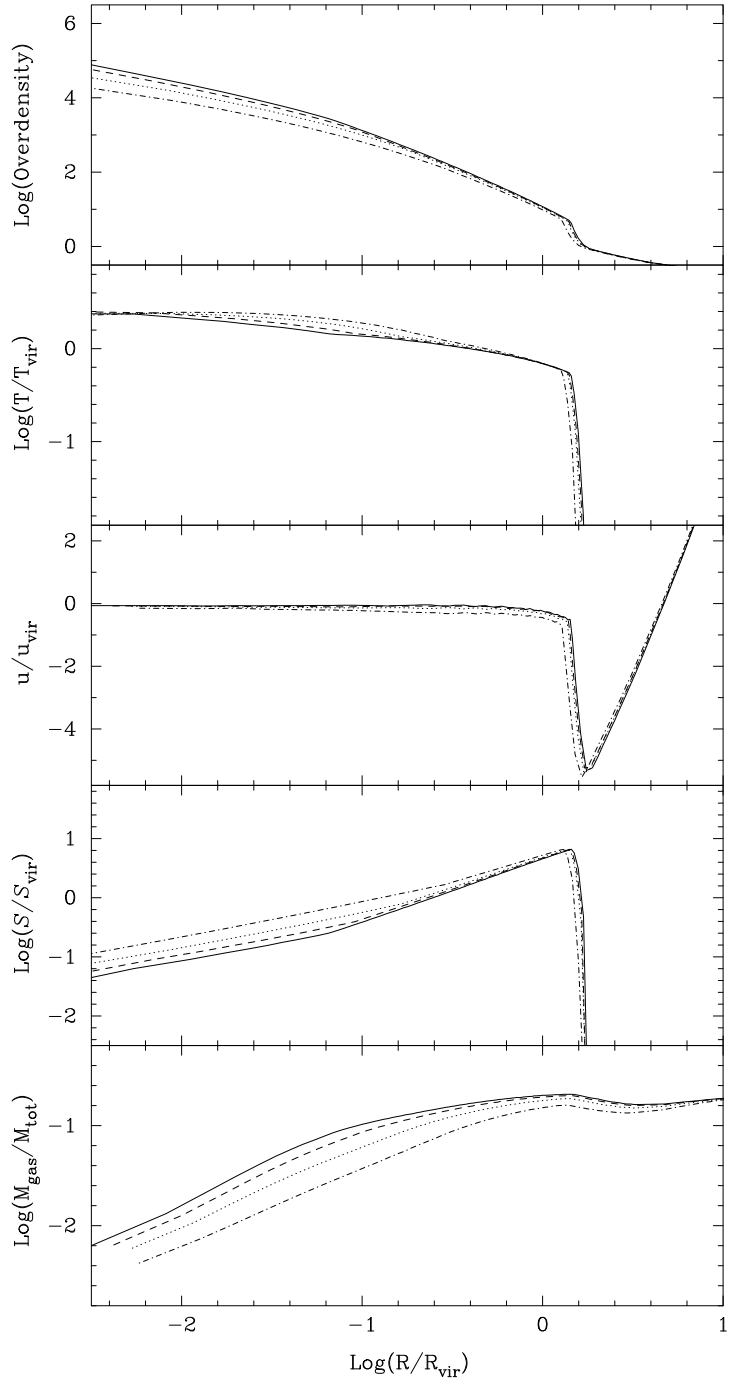}
\caption{Final scaled gas density, temperature, velocity, entropy,
  and enclosed gas mass fraction profiles for standard models of mass 
  $\mv=2.0\times10^{13}$, $7.8\times10^{13}$, $3.9\times10^{14}$ and $1.6\times10^{15}$ 
  $M_{\odot}$ (dot-dash, dotted, dashed, and solid line styles 
  respectively).\label{scale}}
\end{center}
\end{figure}

\fig{scale} shows the final ($z=0$) scaled gas profiles for different mass 
models. The potentials of these clusters have evolved in a self-similar
fashion, as described in section~2.3. If the gas has followed suit, then
the profiles in \fig{scale} should coincide.
At large radius, where there is no additional physics to break the
scaling, the scaled profiles are very similar to each other. However,
significant differences occur near the cluster centre, where cooling is important.

\section{THE COOLING FLOW\label{cf}}

\subsection{Mass Deposition Profile}
\begin{figure}
\begin{center}
\leavevmode
\epsfxsize=15cm
\epsfbox[0 0 543 543]{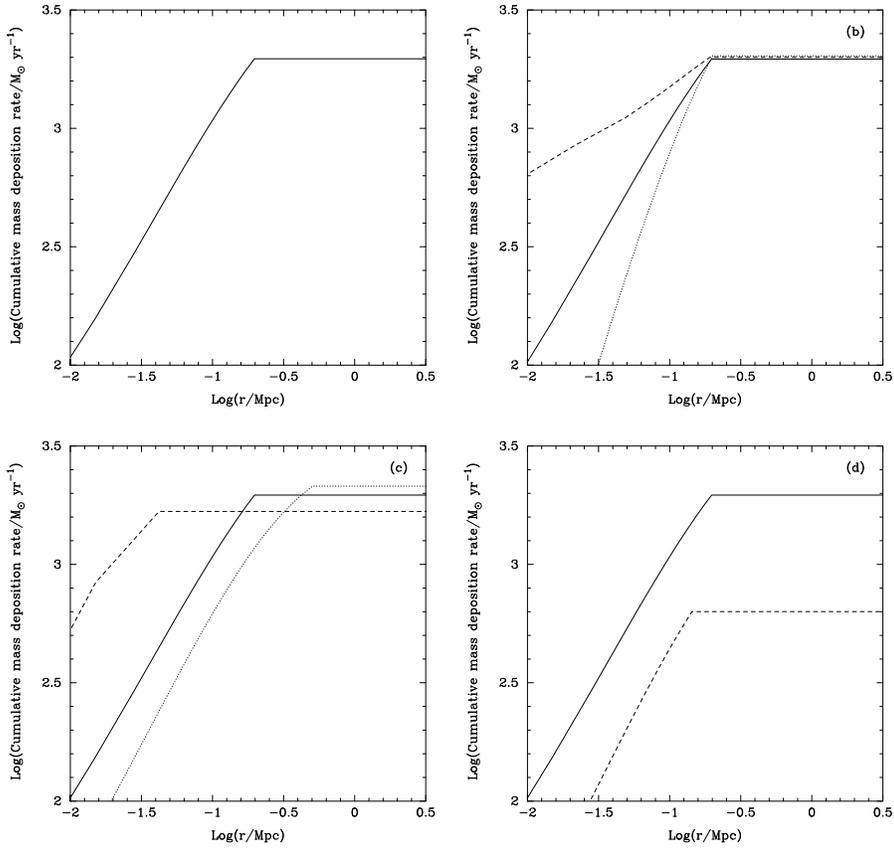}
\caption{Final cumulative mass deposition rate for the $\mv=1.6\times10^{15}$ 
$M_{\odot}$ model. Panels show (a) $\xi=1.7$, $\alpha_{md}=1.0$, 
$f_{gas}=0.2$, (b) $\xi=1.7$, 1.0 and 2.5 (solid, dashed and dotted lines), 
(c) $\alpha_{md}=1.0$, 0.2 and 5.0 (solid, dashed and dotted 
lines), and (d) $f_{gas}=0.2$ and 0.1 (solid and dashed lines
respectively).\label{cf1}}
\end{center}
\end{figure}

As discussed in \sect{describe}, the mass deposition parameter $\xi$ was chosen so
that the mass deposition profile is similar to observations,
$\dot{M}(r)\propto r$ (Fabian, Nulsen \& Canizares 1991). 
The final cumulative mass deposition rate is shown plotted against radius for the 
$\mv=1.6\times10^{15}$ $M_{\odot}$ model in \fig{cf1}a -- within the
mass deposition region $\dot{M}(r)\propto r$, as expected.
\begin{figure}
\begin{center}
\leavevmode
\epsfxsize=9cm
\epsfbox[0 0 543 543]{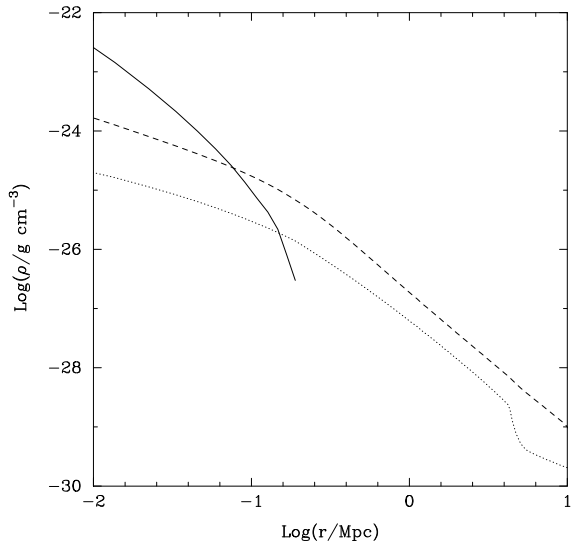}
\caption{Final cooled gas, dark matter and gas density profiles for the 
  $\mv=1.6\times10^{15}$ $M_{\odot}$ model (solid, dashed and dotted lines 
  respectively).\label{cf2}}
\end{center}
\end{figure}
The total mass deposition rate for this model is higher than generally
found observationally -- $\dot{M}\simeq2000$ $M_{\odot}$ compared to
typical observed values of $\lsim1000$ $M_{\odot}$ yr$^{-1}$, although
there are a few exceptional cases with much higher inferred mass
deposition rates (e.g. Schindler \etal\ 1996). The density profile of the
cooled material is centrally peaked and constitutes the dominant
gravitational component at small radii (see \fig{cf2}). The cluster
potential is therefore sufficiently peaked that the cooling flow becomes
gravity dominated near the cluster centre, and the temperature of the hot
phase increases due to adiabatic compression as it flows inwards. As a
result of this dominant cooled gas component in the core, the circular
velocities near the cluster centre are $v_{circ}\simeq1500$~km~s$^{-1}$
in high mass models -- much larger than the velocity
dispersions ($\lsim400$ km s$^{-1}$, Oegerle \& Hoessel 1991)
observed in the envelopes of the cD galaxies which are
commonly found at the focus of cluster cooling flows.
This discrepancy could be reduced by mechanisms not
incorporated in the present models, such as cluster mergers distributing the
material throughout the core, or preheating of the cluster gas, which
reduces the mass of cooled material (see Paper II).

Several possible mechanisms for lowering the mass deposition rate have
been explored: lowering the baryon fraction, $f_{gas}$, artificially
reducing the radiative cooling rate, and varying the model parameters
governing the deposition of gas in the cooling flow ($\xi$ and
$\alpha_{md}$). The final ($z=0$) mass deposition profiles for some of
these models are shown in \fig{cf1}b-d. Except for the models with low
baryon fraction, varying the above parameters has little effect on the
total mass deposition rate. To some extent, there is a feedback mechanism
which keeps the mass deposition rate at an approximately constant value,
regardless of the values specified for the cooling flow parameters -- if
the cooling rate is reduced for example, the gas simply flows further
into the cluster centre (where the density is higher) before dropping out
of the flow. The models with reduced baryon fraction have significantly lower
mass deposition rates, but the gas mass fraction in these cases is in
conflict with observations (White \& Fabian 1995), even if the ICM is
multiphase (Gunn \& Thomas 1995).

Since our attempts at reducing the mass deposition rate by varying the
parameters of the code have been unsuccessful, we conclude that the most
likely explanations for the discrepancy between the mass deposition rate
of the models and values inferred from observations are related to
limitations of the model. For example, (1) the spherical geometry
assumption, and the neglect of angular momentum that would inhibit gas
flowing into the cluster centre, (2) the lack of cluster mergers that
would periodically disrupt the cooling flow, and (3) turbulence and
thermal conduction within the core. Since both angular momentum and
merging would tend to reduce the central density, and hence the mass
deposition rate, we believe that our models represent `maximal' cooling
flows, giving a useful upper bound to mass deposition rates that might be
found observationally. This view receives some support from the observed
variation of mass deposition rate with cluster temperature, as discussed
in the next section.
\begin{figure}
\begin{center}
\leavevmode
\epsfxsize=14cm
\epsfbox[0 0 543 543]{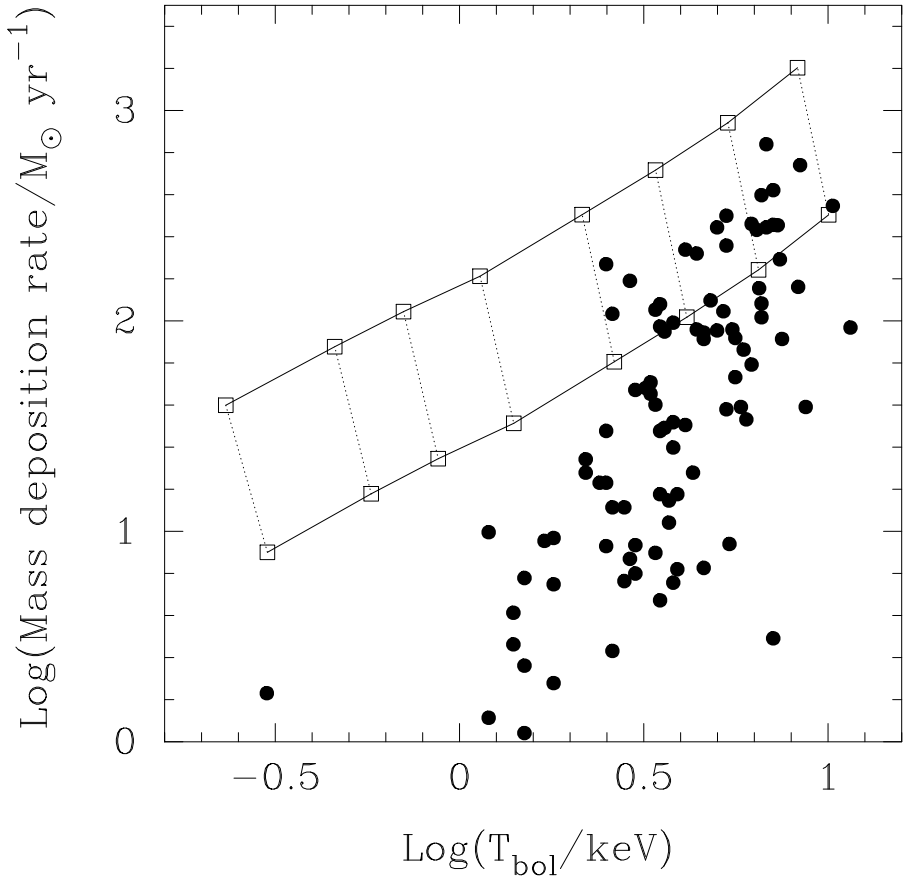}
\caption{Final total mass deposition rate plotted against emission-weighted 
  temperature for models of mass $\mv=7.8\times10^{12}$ $M_{\odot}$, $2.0\times10^{13}$ 
  $M_{\odot}$, $3.9\times10^{13}$ $M_{\odot}$, $7.8\times10^{13}$ $M_{\odot}$, 
  $2.0\times10^{14}$ $M_{\odot}$, $3.9\times10^{14}$ $M_{\odot}$, 
  $7.8\times10^{14}$ $M_{\odot}$ and $1.6\times10^{15}$ $M_{\odot}$. 
  The lower line shows the effect of scaling the mass deposition rate down by
  a factor of five. The observational data are taken from White, Jones \&
Forman (1996).\label{mt1}}
\end{center}
\end{figure}
\begin{figure}
\begin{center}
\leavevmode
\epsfxsize=12cm
\epsfbox[0 0 543 543]{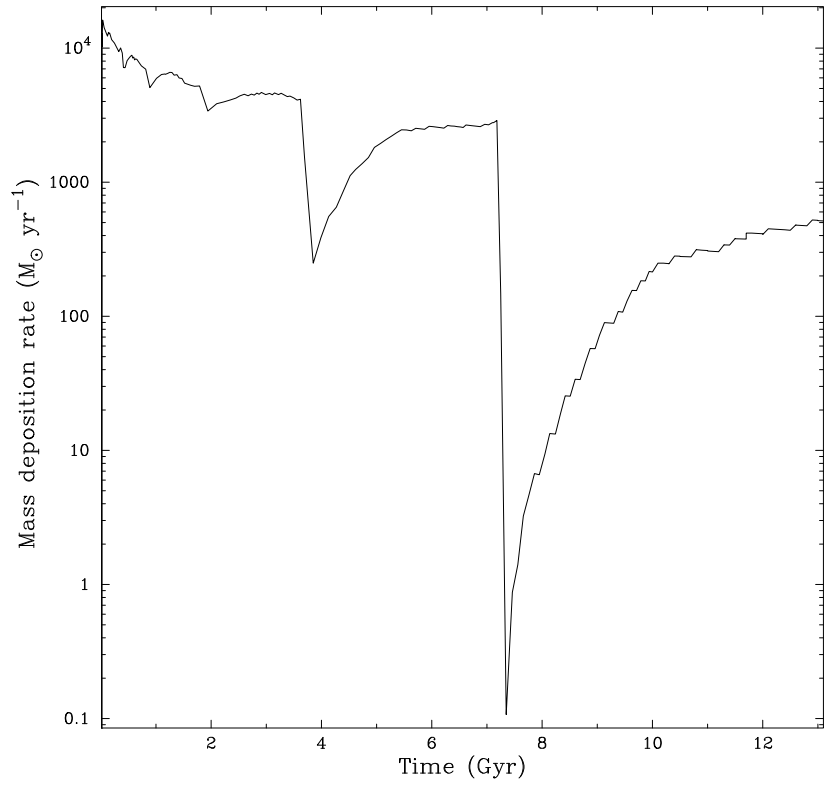}
\caption{Evolution of the total mass deposition rate for a model of mass
  $\mv=1.6\times10^{15}$ $M_{\odot}$ in which the cooling flow is disrupted 
  using the procedure described in the text at several redshifts. \label{lowmdot1}}
\end{center}
\end{figure}
\begin{figure}
\begin{center}
\leavevmode
\epsfxsize=17cm
\epsfbox[0 0 543 543]{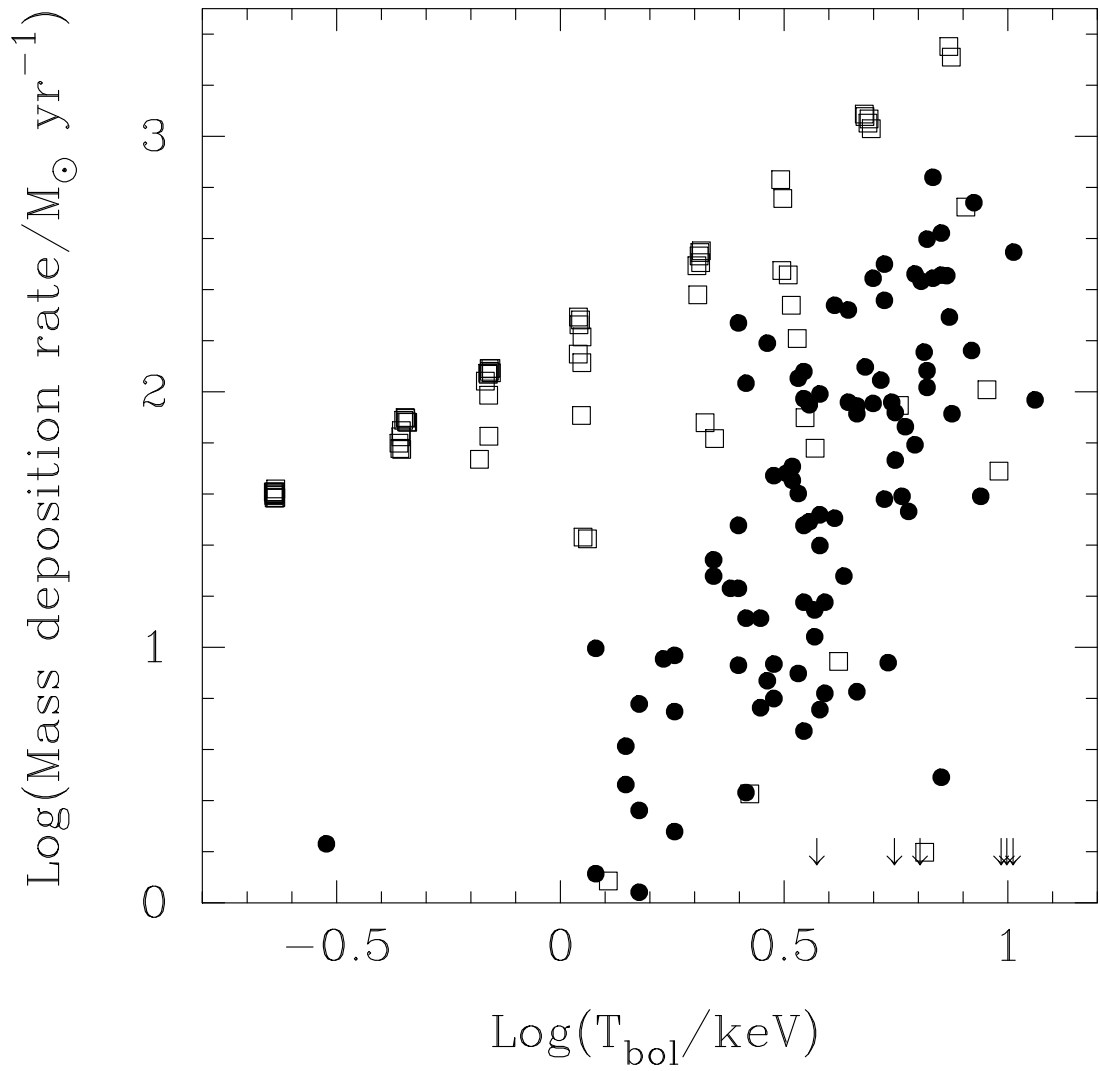}
\caption{Final total mass deposition rate plotted against emission-weighted 
  temperature for the cooling flow disruption models described in the text. 
  The filled circles are observational data taken from White, Jones \&
  Forman (1996), and the
  arrows indicate model clusters with $\dot{M}<1$ $M_{\odot}$ yr$^{-1}$.
  \label{lowmdot2}}
\end{center}
\end{figure}

The high mass deposition rates of the standard models have a knock-on
effect on other quantities, such as the emission-weighted temperature,
and X-ray luminosity. Since the lower mass deposition rates of real
clusters are likely to be due to physical processes not modelled here,
we have shown the effect, in some of the succeeding plots, of
arbitrarily reducing these mass deposition rates by a factor of 5.

\subsection{The Effects of Merging\label{mer}}
\fig{mt1} shows the variation of total mass deposition rate with 
emission-weighted temperature for models of different mass.
Two effects are apparent: (i) the mass deposition rates of the models 
are higher than those found observationally, and (ii) although the
deposition rate increases with mass in the models, the trend is 
less steep than in observed clusters. For high mass systems,
the models shown in \fig{mt1} appear to represent 
an upper bound to the observations, suggesting that
some additional variable (perhaps cluster mergers) interrupts the idealised 
behaviour of the simulations. For example, if cooling flows are disrupted by
major mergers and the time between such mergers is comparable to the
timescale for the cooling flow to re-establish itself following a merger, then
a wide range of mass deposition rates for clusters of a given richness might
be expected.

In order to test this possibility, and also to explore the impact on our
other results, we have performed a series of simulations in which the
cooling flow is disrupted by raising the entropy of the gas
within the cluster core ($r\leq0.2\rv$). The effect of this is
to force gas out of the cluster core, temporarily reducing the mass
deposition rate -- \fig{lowmdot1} shows the evolution of the mass
deposition rate for a simulated rich cluster in which the cooling flow is
disrupted several times during its evolution using this procedure.
To determine whether cooling flow disruption can account for the scatter
in the variation of mass deposition rate with cluster temperature, we have
constructed some models in which the cooling flows are disrupted
periodically, as would be expected in a Universe in which structures
grow largely by hierarchical merging. We assume that clusters typically
experience a major merger each time the Universe doubles its age
(see e.g. Lacey \& Cole 1993; Kauffmann \& White 1993), but we also introduce
some scatter in the merger times, such that
the $n$th major merger occurs {\em on average} at time $t^n$, where
$t^n=2t^{n-1}$, with the {\em actual} merger time being chosen randomly
from the range $(t^{n-1}+t^n)/2\rightarrow(t^n+t^{n+1})/2$. The results
from 80 such models are shown in \fig{lowmdot2}.

Clearly, disruption of the cooling flow can substantially reduce the
present day mass deposition rates for the more massive systems since, for
these systems, the timescale for the cooling flow to re-establish itself
is comparable to the time between mergers. However, the low mass systems
have substantially lower cooling times in their cores and the cooling
flow is able to re-establish itself much faster, hence they are, on
average, little affected by cooling flow disruption. For massive systems,
the scatter in the mass deposition rate for a given temperature appears
to be similar to that observed, but high resolution
3-D simulations with radiative cooling are needed to definitively
establish the effects of cluster merging on cooling flows.

\subsection{Evolution of the Cooling Flow}
\begin{figure}
\begin{center}
\leavevmode
\epsfxsize=14cm
\epsfbox[0 0 543 543]{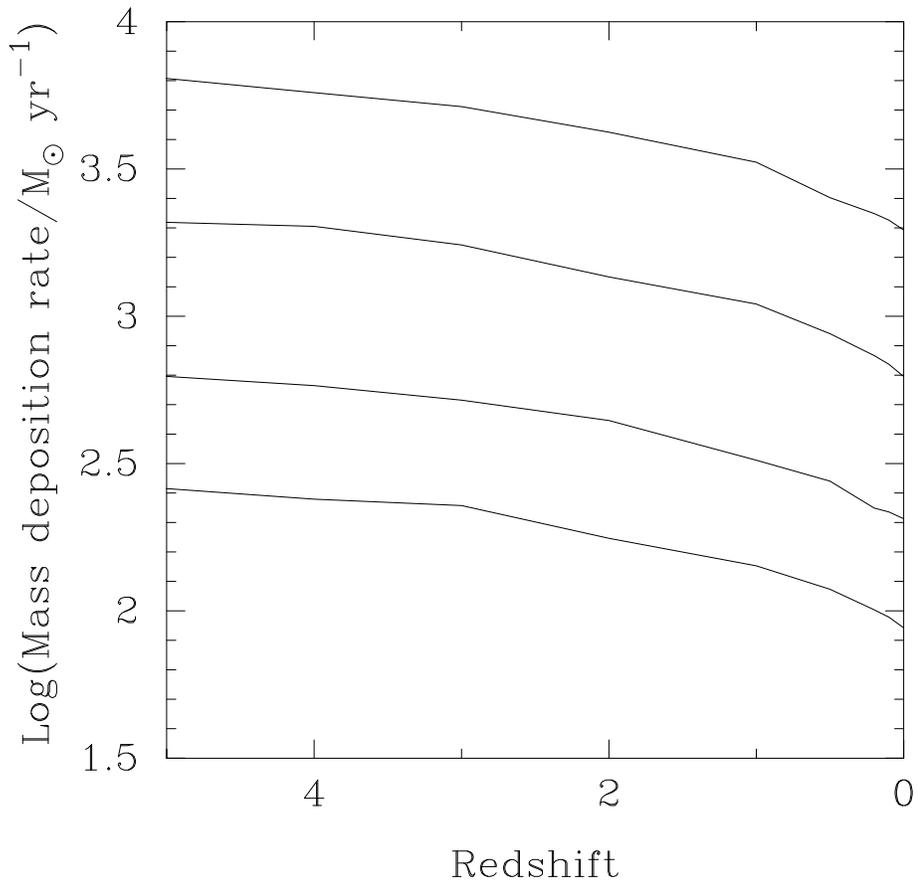}
\caption{Evolution of the mass deposition rate for models with 
  (from top to bottom) $\mv=1.6\times10^{15}$ $M_{\odot}$, 
  $3.9\times10^{14}$ $M_{\odot}$, $7.8\times10^{13}$ $M_{\odot}$, 
  and $2.0\times10^{13}$ $M_{\odot}$.\label{mz1}}
\end{center}
\end{figure}

The evolution of the mass deposition rate is shown for a range of system
masses in \fig{mz1}. The mass deposition rate is substantially (a factor
of 2--3) larger at $z=5$ than at $z=0$, decreasing
with time due to the inflow of higher entropy gas from larger
radii to replace that deposited at the cluster centre. This lowers the
central gas density, which in turn reduces the mass deposition rate.
As a result, most deposition occurs at high redshift, with $\sim55$\% of the total
amount dropped by $z\simeq1$.

\section{THE GAS DISTRIBUTION\label{fgas}}

\subsection{The X-ray Surface Brightness Profile\label{sbprof}}
\begin{figure}
\begin{center}
\leavevmode
\epsfxsize=14cm
\epsfbox[0 0 464 454]{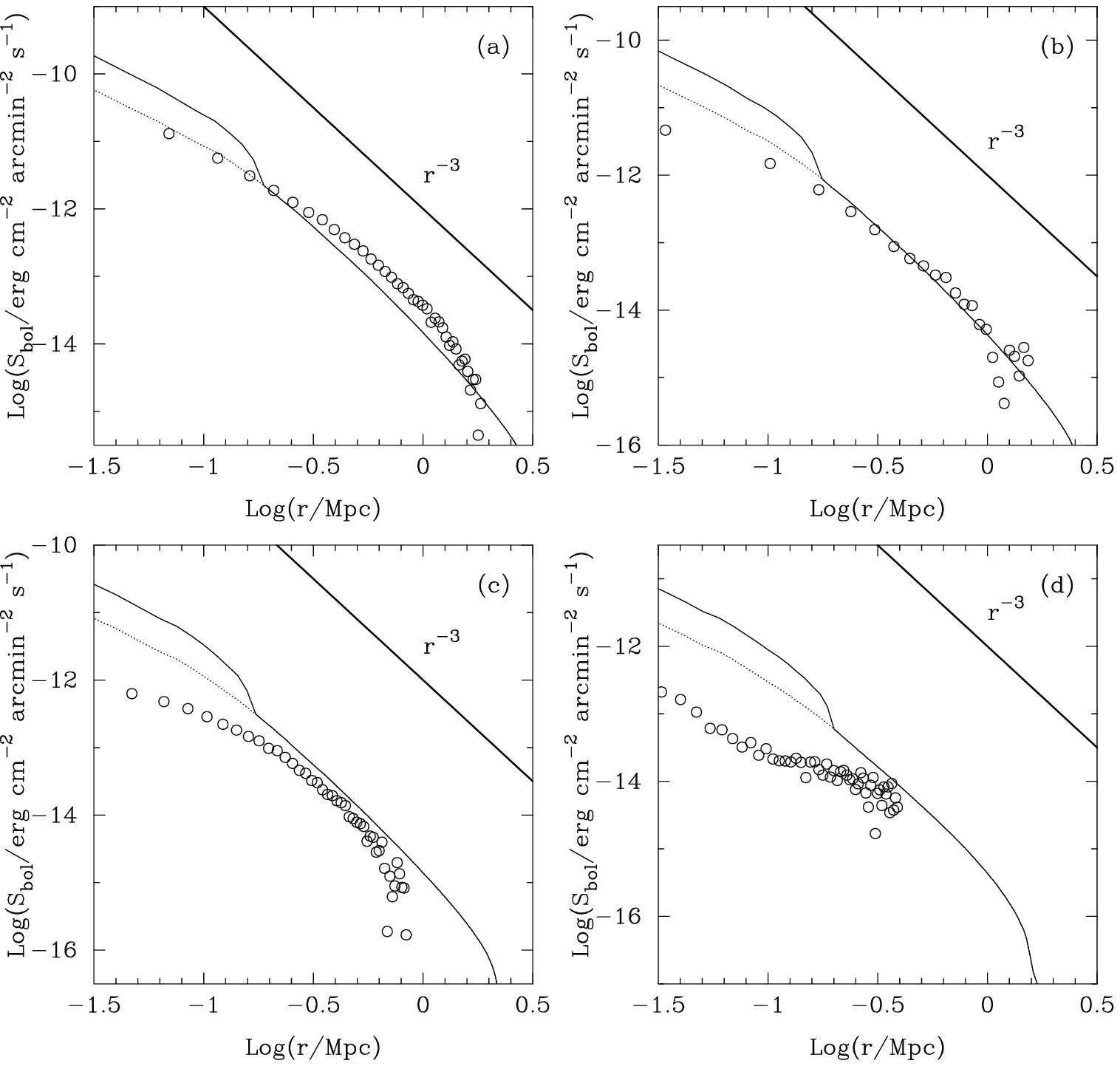}
\caption{Final bolometric surface brightness profiles for systems of mass
  (a) $\mv=7.8\times10^{14}$
  $M_{\odot}$, (b) $\mv=3.9\times10^{14}$ $M_{\odot}$, (c) $\mv=2.0\times10^{14}$
  $M_{\odot}$ and (d) $7.8\times10^{13}$ $M_{\odot}$. The dotted line shows the
  surface brightness when the mass deposition luminosity is ignored.
  The circles indicate the 
  surface brightness profiles of the following clusters observed by the 
  {\em ROSAT} PSPC (D. Cannon, private communication; Ponman \& Bertram 1993): 
  (a) A1795, (b) MKW3, (c) A1060, and (d) Hickson 62. Note that the models have
  not been optimised to match the data, which are illustrative only. 
  The relationship $S\propto r^{-3}$ is shown for reference, and 
  $H_0=50$ km s$^{-1}$ Mpc$^{-1}$ is assumed.\label{surf1}}
\end{center}
\end{figure}

Final $z=0$ bolometric surface brightness profiles for different mass
models are shown in \fig{surf1}, together with observed profiles for
groups and clusters of comparable mass (\tab{tclus}). 
The observational profiles are based on \ROSAT\ PSPC data converted to
bolometric flux using \ROSAT\ or \Ginga\ temperatures, assuming
isothermal gas (Cannon, Ponman \& Navarro, in preparation). 
The model profiles were calculated by
integrating the emission from the shells along each line of sight, using
the cooling function of Raymond, Cox \& Smith (1976). An additional 
component due to gas cooling to zero temperature at a rate proportional
to the local mass deposition rate was also included.
\begin{table}
\begin{center}
\begin{tabular}{lccc}\hline\hline
\multicolumn{1}{c}{Cluster} & Mean cluster    & Final virial mass & Final mean model   \\
           & temperature$^{\dagger}$ & of model       & temperature        \\
           & $\overline{T}$/keV  & $\mv/M_{\odot}$& $\overline{T}$/keV \\ \hline
Hickson 62 & 0.96          & $7.8\times10^{13}$  & 1.1 \\
A1060      & 2.55          & $2.0\times10^{14}$  & 2.2 \\
MKW 3      & 3.0          & $3.9\times10^{14}$  & 3.4 \\
A1795      & 5.34          & $7.8\times10^{14}$  & 5.4 \\ \hline
\multicolumn{4}{l}{$\dagger$ Ponman \& Bertram 1993; Yamashita 1992}
\end{tabular}
\caption{Mean temperatures for the clusters in Figure 12a--d.\label{tclus}}
\end{center}
\end{table}

The surface brightness profiles of the more massive systems are similar
to, but slightly steeper than, those of observed clusters of similar
mass: at the centre, the very large contribution of mass deposition to
the cluster luminosity is clearly visible as a region of excess emission.
Beyond this region, the surface brightness profile gradually steepens
with radius. The shape of the {\em EVOL} surface brightness profiles of
the richer systems are quite similar to the observed profiles outside the
central region. However, unlike the {\em EVOL} profiles, there is no
evidence for a region of greatly enhanced emission due to mass deposition
within the cooling flow near the centre of observed clusters (though the
isothermal assumption may in practice lead to an underestimate of the
bolometric surface brightness by $\sim20$\% within the cooling flow
region). The mismatch between the models and data near the centre
persists even if the numerical parameters in {\em EVOL} are adjusted to
confine the mass deposition to a region closer to the cluster centre. The
discrepancy could be resolved if the cooling flows of real clusters are
periodically disrupted (\sect{cf}). As can be seen from the figure, the
model profiles of the massive systems match the data much better if the
emission from the model cooling flow is omitted.

A fit of the function
\begin{equation}
  S(r)=S(0)\left[1+\left(\frac{r}{r_c}\right)^2\right]^{-3\beta_{fit}+1/2},
  \label{surfb}
\end{equation}
to the underlying cluster emission (\ie\ {\em excluding} the additional
luminosity due to mass deposition) between $r=0$ and $0.5\rv$ yields a
$\beta_{fit}$ value of $\sim$0.72. This is slightly larger than inferred
from observations of rich clusters, which generally yield
$<\!\!\beta_{fit}\!\!>\simeq0.67$ and hence $S_x\propto r^{-3}$ 
outside the core (Jones \& Forman 1984). 

For lower mass clusters, the emission from
the models falls above that found observationally, with the largest
discrepancy being found in the smallest systems. This suggests that the
gas fraction in the inner regions declines from rich clusters to groups,
as found by David, Jones \& Forman (1995).

\subsection{$\beta_{fit}$ vs Temperature and the Gas Entropy}
\begin{figure}
\begin{center}
\leavevmode
\epsfxsize=16cm
\epsfbox[0 0 543 543]{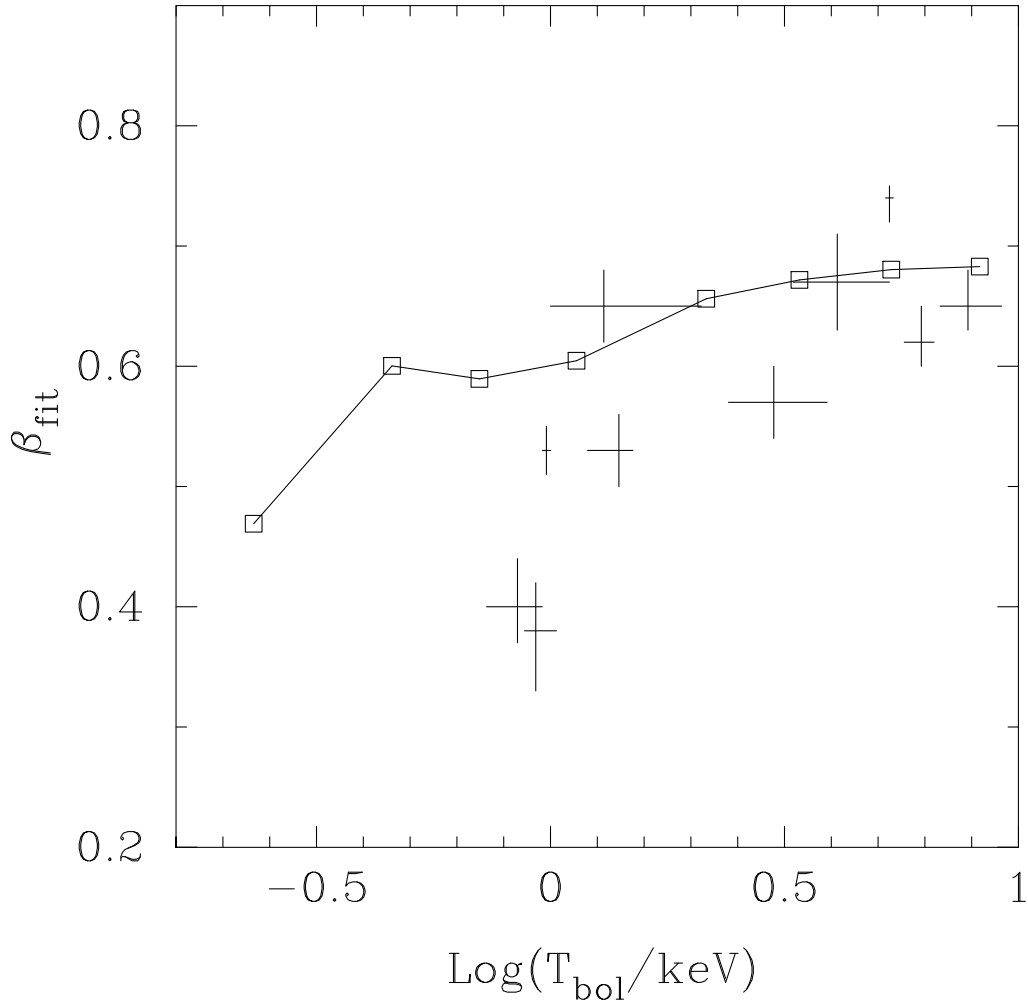}
\caption{Final $\beta_{fit}$ plotted against emission-weighted temperature for
  models of mass $\mv=2.0\times10^{13}$ $M_{\odot}$, $3.9\times10^{13}$ $M_{\odot}$, 
  $7.8\times10^{13}$ $M_{\odot}$, $2.0\times10^{14}$ $M_{\odot}$, $3.9\times10^{14}$ 
  $M_{\odot}$, $7.8\times10^{14}$ $M_{\odot}$ and $1.6\times10^{15}$ $M_{\odot}$. 
  The observational data (crosses) are taken from David, Jones \& 
  Forman (1995).\label{beta1}}
\end{center}
\end{figure}

Figure~\ref{beta1} shows the variation of $\beta_{fit}$ 
with emission-weighted temperature at $z=0$ for models of different
mass. 
Also shown are values of $\beta_{fit}$ from an observed sample
(David, Jones \& Forman 1995). 

The results confirm those of \sect{sbprof} -- there is a modest trend for
a decrease in $\beta_{fit}$ with decreasing $\overline{T}$, although
$\beta_{fit}$ is found to drop substantially for the lowest mass system,
probably due to the importance of line cooling. Reasonable agreement with
observations is found for the rich systems which have
$\beta_{fit}\simeq0.68$, very similar to the values found observationally
($\beta_{fit}\simeq0.67$). There is, however, a large discrepency between
the simulated clusters and observations for the poorer clusters.
Observationally, $\beta_{fit}$ is found to decline steeply with
decreasing $T$, which suggests the presence of an additional mechanism
operating in poorer systems that has not been included in the simulations
presented here.

Another pointer to the need for additional physical processes in low mass
systems is the value of the entropy of the intracluster gas. Gas in the
inner regions of X-ray bright galaxy groups (outside any cooling flow) is
observed to have an entropy corresponding to $S=T/n^{2\over3}\sim
200-300$~keV~cm$^2$ (Bourner and Ponman, paper in preparation) -- very
similar to that found in cluster cores. In the {\em EVOL} models presented
here, the gas entropy at the corresponding point in rich systems has
$S\sim200-400$~keV~cm$^2$, as observed, but in the low mass models
($T\sim 1$~keV) it has $S\lsim 50$~keV~cm$^2$.

\subsection{The Gas Mass Fraction Profile}
\begin{figure}
\begin{center}
\leavevmode
\epsfxsize=15cm
\epsfbox[0 0 543 543]{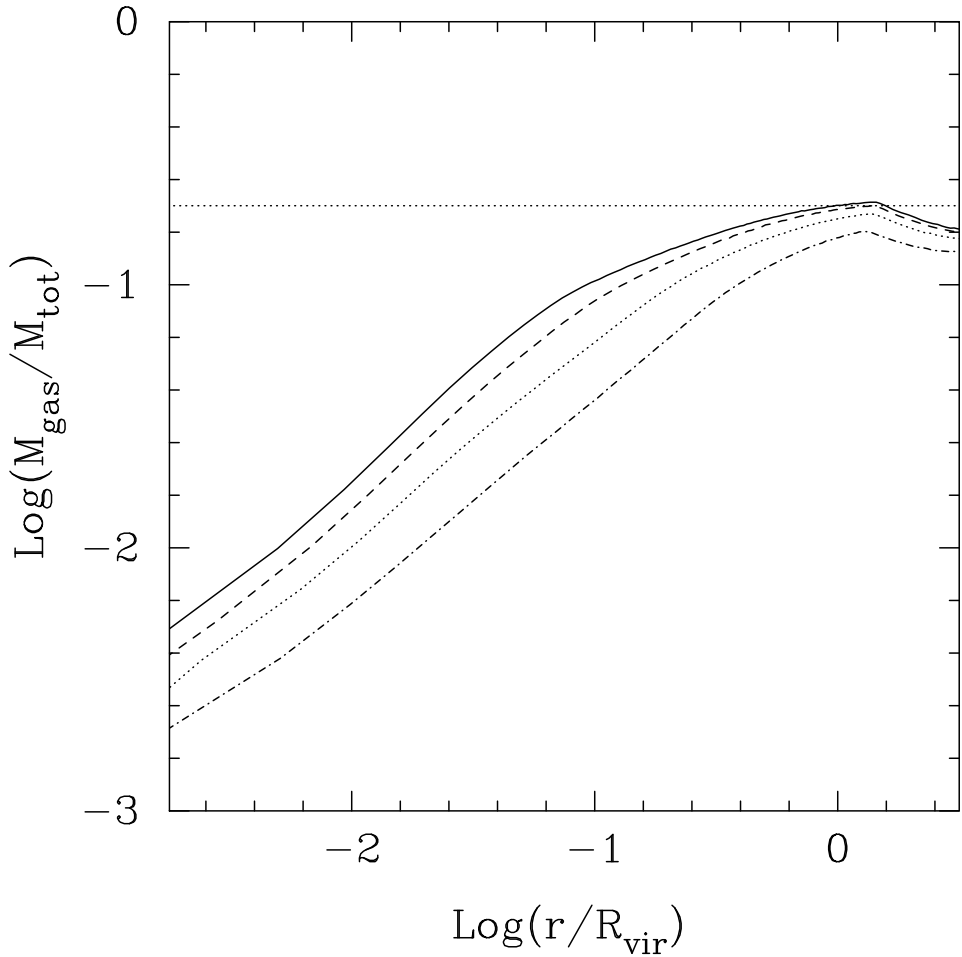}
\caption{Final enclosed gas mass fraction profiles for models of mass 
  $\mv=1.6\times10^{15}$ $M_{\odot}$, $3.9\times10^{14}$ $M_{\odot}$,  
  $7.8\times10^{13}$ $M_{\odot}$ and $2.0\times10^{13}$ $M_{\odot}$
  (solid, dashed, dotted, dot-dash lines respectively). The
  horizontal dotted line shows the global value.\label{frac1}}
\end{center}
\end{figure}

Figure~\myref{frac1} shows the final enclosed gas mass fraction profiles for 
different mass models. In each case, the processes of cooling and mass 
deposition, particularly at early epochs, lead to the development of a profile
that rises continually with radius (gas which has cooled out is
counted as mass, but not as `gas'). Such a rise is similar to that inferred 
by White \& Fabian (1995) from a sample of Abell clusters, and also agrees
reasonably well with the results of David, Jones \& Forman (1995). 
The mechanism behind this is 
interesting -- since the entropy of the gas increases with radius, mass
deposition near the cluster centre leads to the inflow of higher entropy
gas from further out in the cluster which, in turn, lowers the central
gas density producing a gas ``core''. The {\em total} mass density
profile is, however, quite sharply peaked despite the core in the dark
matter profile, being dominated by ``cooled'' gas at small radii
(\fig{cf2}). The net result is a gas mass fraction profile that increases
with radius. The slight decline in $f_{gas}$ seen outside \rv\ arises from
the fact that the gas density drops sharply at an outward propagating
shock, whilst the dark matter density (which is represented by a cored
power law model, as discussed in section 2.3) declines smoothly to 
the background value.

Although central cooling and consequent gas inflow does produce a rising
gas fraction profile, in accord with observations, it involves mass
deposition rates, particularly at large redshift, which are much larger
than those typically inferred for nearby clusters. Alternatively, there
are several possible mechanisms, such as major mergers, and energy
injection from the cluster galaxies, which may be effective at forcing
gas out of the cluster core -- this would reduce the gas fraction in the
inner regions, and should also lower the mass deposition rate in the
cooling flow.

Figure~\myref{frac1} also shows that $f_{gas}$ in the inner regions
declines somewhat with decreasing system size, due to the greater
importance of cooling in lower mass systems.

\subsection{The Effect of a Fully Multiphase ICM}
\begin{figure}
\begin{center}
\leavevmode
\epsfxsize=14cm
\epsfbox[0 0 464 454]{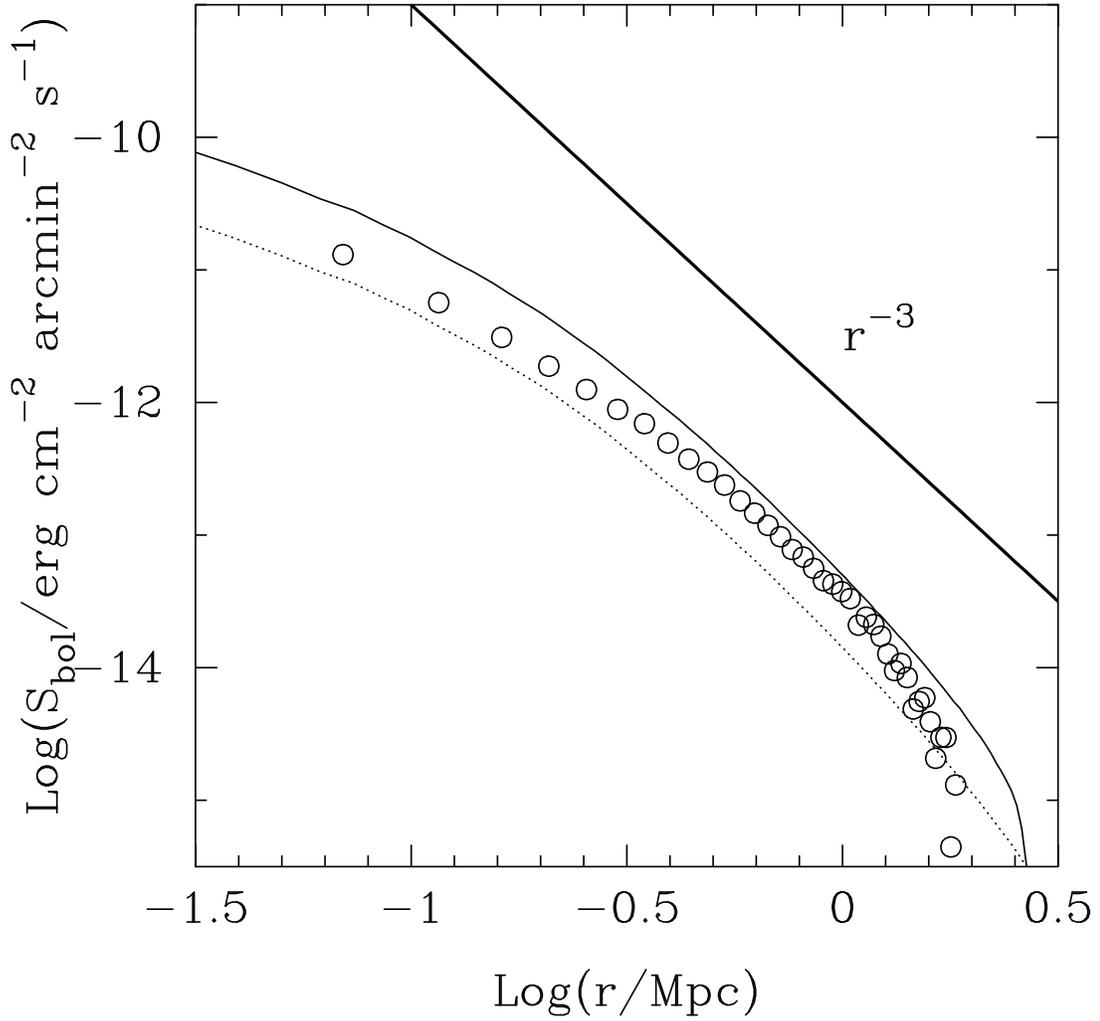}
\caption{Final bolometric surface brightness profiles for a fully multiphase ICM model
  of mass $7.8\times10^{14}$ $M_{\odot}$. The dotted line shows the
  surface brightness when the mass deposition luminosity is ignored.
  The circles indicate the surface brightness profile of A1795 observed by the 
  {\em ROSAT} PSPC (D. Cannon, private communication).\label{surf2}}
\end{center}
\end{figure}

Since the cooling time threshold for the onset of multiphase mass
deposition is somewhat arbitrary, we have examined the alternative
possibility that the bulk of the ICM is multiphase by constructing some 
models in which mass deposition is allowed to
occur throughout the cluster ($\alpha_{md}=\infty$). The final ($z=0$) surface 
brightness profile for such a model of mass $\mv=7.8\times10^{14}$ $M_{\odot}$
is shown in \fig{surf2}. 
The shape of the surface brightness profile is similar to that of
observed clusters, but the contribution of the cooling gas to the cluster
luminosity is appreciable at all radii ($\sim70$\% of the total
luminosity). This reduces the gas fraction required for a given cluster
luminosity by $\sim70$\% and would be a possible mechanism for
alleviating the ``baryon catastrophe'' associated with clusters of
galaxies (Gunn \& Thomas 1995). However, it was shown some time ago by
Allen \etal\ (1992) that the spectral properties of the intracluster
emission in the Perseus cluster are inconsistent with such
large scale multiphase structure, and this conclusion has been confirmed
for other systems by ASCA (Mushotzky \etal\ 1996).

\section{X-RAY LUMINOSITY\label{lum}}

\subsection{Luminosity-Temperature Relation}
\begin{figure}
\begin{center}
\leavevmode
\epsfxsize=14cm
\epsfbox[0 0 543 543]{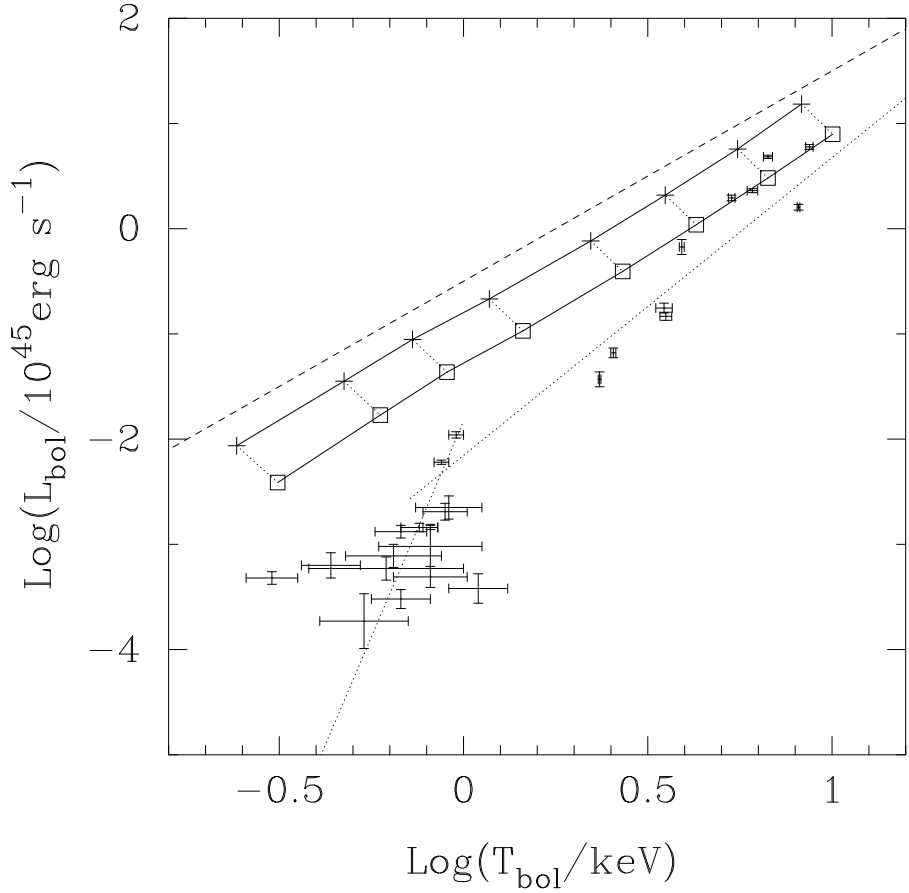}
\caption{Bolometric luminosity plotted against temperature for models
  of mass $\mv=7.8\times10^{12}$ $M_{\odot}$, $2.0\times10^{13}$ 
  $M_{\odot}$, $3.9\times10^{14}$ $M_{\odot}$,
  $7.8\times10^{13}$ $M_{\odot}$, $2.0\times10^{14}$ $M_{\odot}$,
  $3.9\times10^{14}$ $M_{\odot}$, $7.8\times10^{14}$ $M_{\odot}$ and
  $1.6\times10^{15}$ $M_{\odot}$. The lower line shows the effect of
  scaling the mass deposition rate down by a factor of 5. 
  The data points are from Ponman \etal\ (1996), and the dashed line
  (which has arbitrary normalisation) shows the relation, $L\propto T^2$,
  expected for self-similar cluster structure.\label{lt1}}
\end{center}
\end{figure}

Figure~\ref{lt1} shows the bolometric luminosity, $L_{bol}$, plotted 
against emission-weighted temperature, $T_{bol}$, for various models.
The calculation of $T_{bol}$ includes the contribution
of mass deposition within the cooling flow. For the hottest models,
this ``cooling luminosity'' lowers the emission-weighted temperature 
by up to $\sim$30\%. 
\begin{figure}
\begin{center}
\leavevmode
\epsfxsize=14cm
\epsfbox[0 0 543 543]{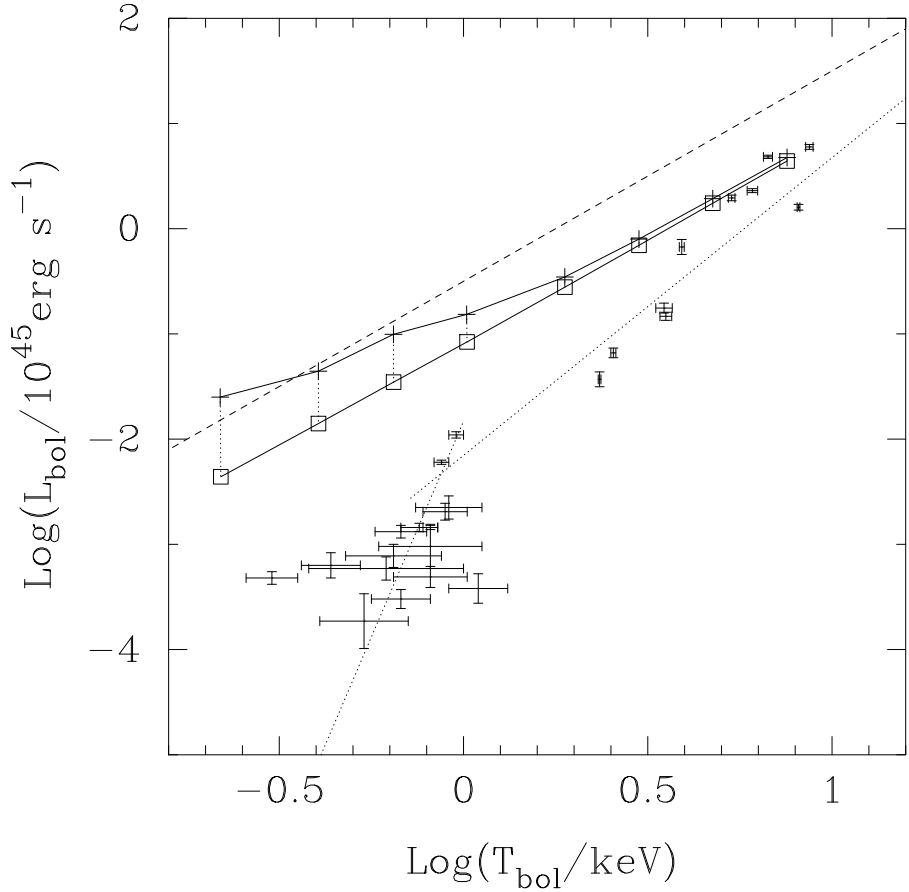}
\caption{Bolometric luminosity plotted against temperature for models
  in which radiative cooling has been turned off during the evolution.
  The masses of the clusters are
  $\mv=7.8\times10^{12}$ $M_{\odot}$, $2.0\times10^{13}$ 
  $M_{\odot}$, $3.9\times10^{14}$ $M_{\odot}$,
  $7.8\times10^{13}$ $M_{\odot}$, $2.0\times10^{14}$ $M_{\odot}$,
  $3.9\times10^{14}$ $M_{\odot}$, $7.8\times10^{14}$ $M_{\odot}$ and
  $1.6\times10^{15}$ $M_{\odot}$. The lower line shows the effect of
  setting the metal abundance to zero when calculating the luminosity and
  mean temperature.
  The data points are from Ponman \etal\ (1996), and the dashed line
  (which has arbitrary normalisation) is the prediction for self-similar 
  cluster structure, $L\propto T^2$.\label{lt2}}
\end{center}
\end{figure}
If the structure of all clusters is self-similar 
then scaling relations can be derived for the evolution of mean cluster
properties. One of the predictions of self-similarity is a relationship
between bolometric luminosity and temperature of the form
$L_{bol}\propto T^2$ (NFW96), much flatter than the observed relationship,
$L_{bol}\propto T^{2.81\pm0.18}$ (White, Jones \& Forman 1996).

The slope of the $L:T$ relation for the infall models considered here is
2.1 (2.2 if the mass deposition rate is reduced by a factor of five),
which is still flatter than that found observationally. In order to
investigate the effects of cooling on the lower mass systems, some
simulations were performed in which radiative cooling was turned off
during the cluster evolution. The results of viewing the X-ray emission
from such a model at $z=0$ are shown in
\fig{lt2}. (The surface brightness profiles of these models are
so steep that the luminosity diverges at small radii so, for these models
only, emission within a radius of $0.05\rv$ has been ignored.) The
greater importance of line cooling to the bolometric luminosity for the
low mass systems causes the $L:T$ relation to flatten at low
temperatures. However, when line radiation is neglected, the logarithmic
slope of the $L:T$ relation is 1.95, very close to the value of 2
expected if the cluster population is self-similar, with different mass
clusters having profiles of the same form.

This test shows that the fact that the slope of the $L:T$ relation in
\fig{lt1} is so similar to that expected for self-similar scaling is
somewhat fortuitous. It occurs because the flattening in the relation due
to the increased contribution of line emission and mass deposition in low
mass systems is compensated by the fact that this extra mass
deposition lowers the gas fraction, which steepens the $L:T$ relation.
The net result is an $L:T$ index only slightly steeper than 2.

The second effect which is apparent in \fig{lt1} is that the
normalisation of the luminosity of the simulated clusters is larger than
that observed. This discrepancy is partly due to the fact that
the mass deposition rates are considerably 
larger than typically inferred from observations. The contribution of
this cooling material to the overall cluster luminosity is significant
(see
\fig{surf1}) -- this is a generic feature of the simple infall models,
possibly compounded by the simplified radial cooling flow.

\begin{figure}
\begin{center}
\leavevmode
\epsfxsize=14cm
\epsfbox[0 0 543 543]{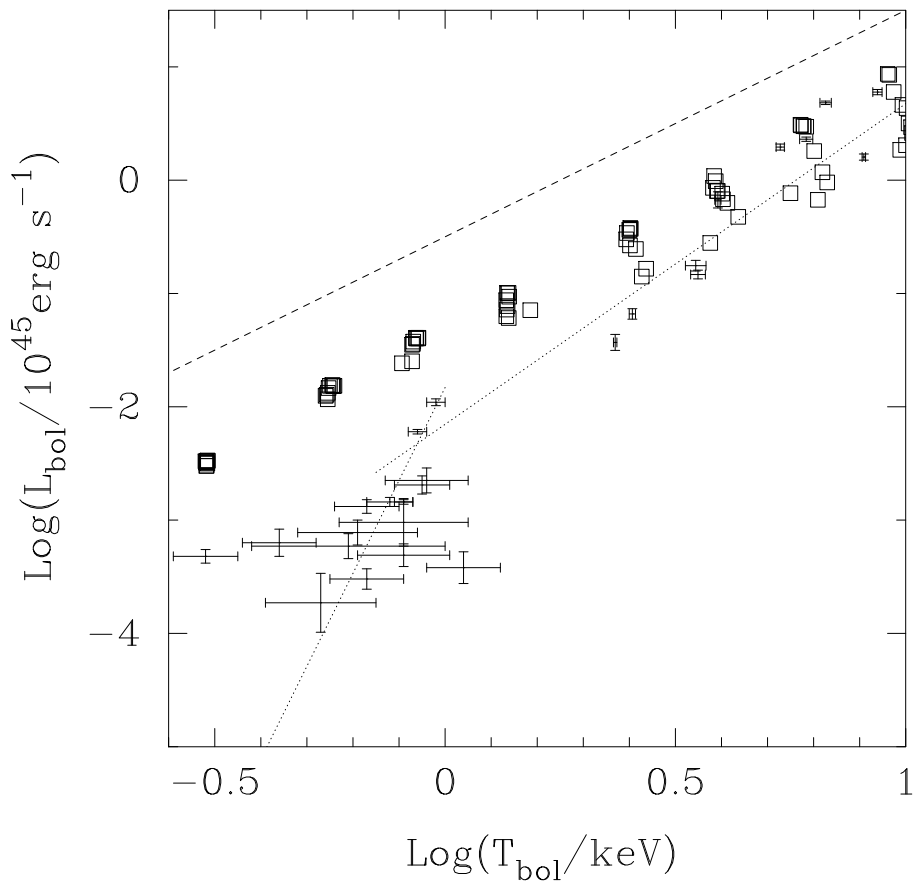}
\caption{Final bolometric luminosity plotted against emission-weighted
  temperature for cooling flow disruption models.
  The data points are from Ponman \etal\ (1996), and the dashed line
  (which has arbitrary normalisation) shows the relation, $L\propto T^2$,
  expected for self-similar cluster structure.\label{merlt}}
\end{center}
\end{figure}

In \fig{merlt} we show the $L:T$ relation for the set of merger models
discussed in \sect{mer}. The effects of merging are to substantially reduce
the luminosity of the more massive ($T\gsim3$~keV) systems, bringing them
into line with observed clusters. However, as noted in \sect{mer}, the
effect of merging on lower mass systems is much less marked, due to
the speed with which the system recovers from a merger, and re-establishes
its cooling flow.

\subsection{Luminosity Evolution}
\begin{figure}
\begin{center}
\leavevmode
\epsfxsize=14cm
\epsfbox[0 0 543 543]{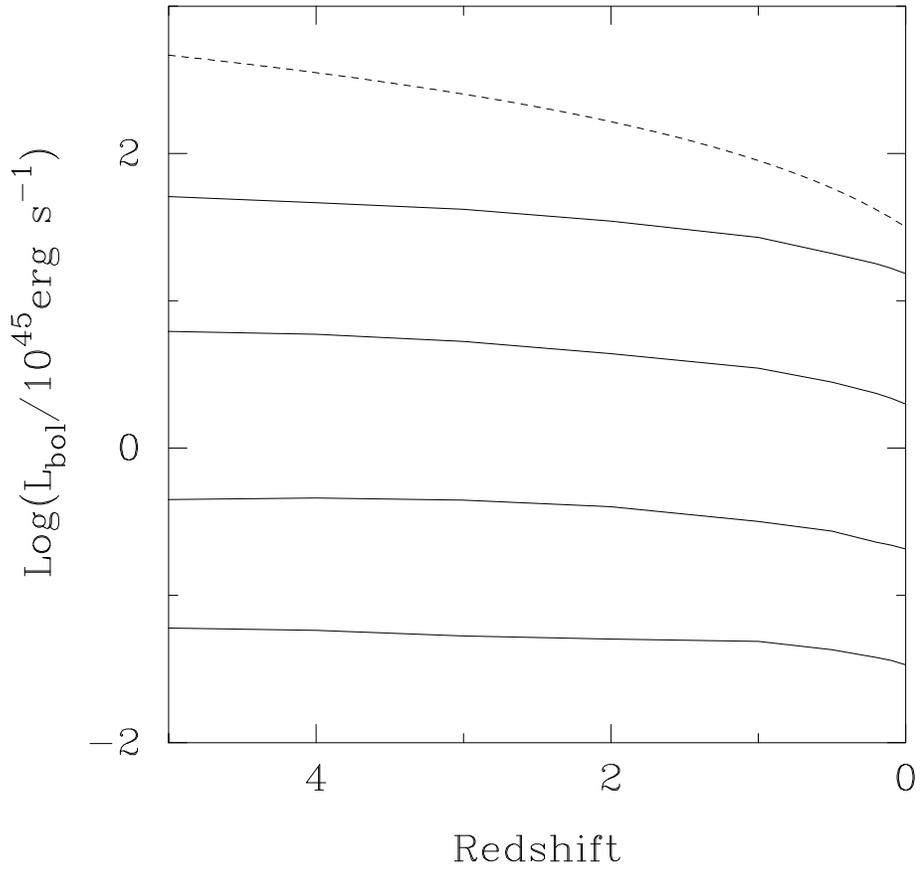}
\caption{Evolution of the bolometric luminosity for models of mass 
  (from top to bottom) $\mv=1.6\times10^{15}$ $M_{\odot}$, 
  $3.9\times10^{14}$ $M_{\odot}$, $7.8\times10^{13}$ $M_{\odot}$, and
  $2.0\times10^{13}$ $M_{\odot}$. The dashed line shows the self-similar 
  prediction.\label{lz1}}
\end{center}
\end{figure}

The evolution of the bolometric luminosity for models with a range of
masses is shown in \fig{lz1}. X-ray emission is generally
detectable out to 0.2--0.5 Mpc and $\sim$2 Mpc for groups and clusters
respectively (Mulchaey \etal\ 1996; Ponman \etal\ 1996; White \& Fabian
1995), corresponding to
a radius of $\sim0.5\rv$. Accordingly, the luminosities given here were
calculated by summing the emission from shells within a radius of
$0.5\rv$.
A line showing the evolution expected if the gas distribution is self-similar
(\eg\ if $\rho_{gas}(r)\propto\rho_{dark}(r)$) is also shown in \fig{lz1},
\begin{equation}
  L_{bol} \propto (1+z)^{(13+7n)/(6+2n)}.\label{lexp}
\end{equation}
This relation is found to give a very good match to the evolution of the 
models with no radiative cooling. 
However, as shown in the figure, models with radiative cooling evolve much less
strongly -- the cluster luminosity rises by $\sim$50\% out to $z=0.5$, while
\eqn{lexp} gives a rise of $\sim$80\% for $n=-1$ over the same interval.
The difference is predominantly due to evolution of the central gas mass
fraction: at earlier epochs, the characteristic density of the virialised
region is higher and mass deposition more widespread, resulting in a
lower gas mass fraction within any fixed fraction of the virial radius
(see \fig{inf2}). As a
result, the gas mass fraction rises with time compared to the
self-similar scaling, reducing the evolution of the luminosity. Such
modest ($\sim$30\% out to $z=0.3$) positive evolution of the X-ray
luminosity of rich clusters is consistent with recent results from
clusters observed during the {\em ROSAT} All Sky Survey (H. Ebeling,
private communication).

\section{THE EFFECT OF $\Omega_0<1$\label{omega}}
In order to test the extent to which the results are sensitive to the
assumed value of $\Omega_0$, a few models have been considered in which
$\Omega_0<1$. In this case, the evolution of the dark matter density
departs significantly from the self-similar behaviour at redshifts
$z\lsim z_{crit}=\Omega_0^{-1}-1$ (Lacey \& Cole 1993). To simulate the
approximate evolution of a cluster in an open Universe, it is assumed 
that the dark matter density behaves exactly
as in the $\Omega=1$ case for $z>z_{crit}$, whilst for $z<z_{crit}$
the dark matter density beyond a radius of $r_{crit}=5\rv$ 
(roughly the radius at which material is just turning around from the 
Hubble expansion) is constant, and equal to the mean density of the 
universe at that epoch.

The treatment of the dark matter within $r_{crit}$ after $z=z_{crit}$ is
somewhat problematic, due to uncertainty in the evolution of the dark
matter ``core''. We consider two cases: (1) at $z=z_{crit}$, the profile
within $r_{crit}$ is frozen and there is no subsequent evolution, and (2)
the dark matter profile within $r_{crit}$ is allowed to evolve in the
same way as in the critical density case. Clearly the behaviour of the
low $\Omega_0$ models in both cases will be unrealistic at large radius
-- the density at $r_{crit}$ will suddenly drop into the background,
whereas in reality one would expect a gradual steepening of the density
profile at large radius (Hoffman \& Shaham 1985; Crone, Evrard \&
Richstone 1994). However, the dynamics in the central region, which
contributes most of the luminosity, will be relatively unaffected by
this.

\begin{figure}
\begin{center}
\leavevmode
\epsfxsize=14cm
\epsfbox[0 0 543 543]{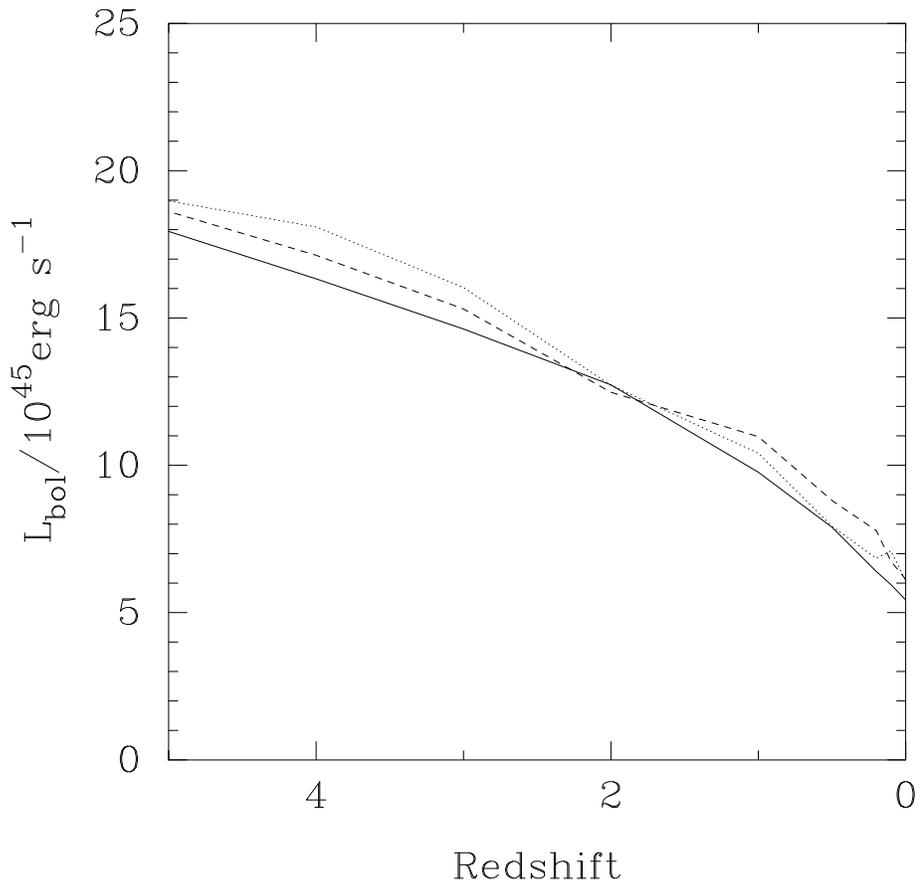}
\caption{Evolution of the bolometric luminosity for models with
  (1) $\Omega_0=1$ and $\mv=1.6\times10^{15}$ $M_{\odot}$ (solid line), 
  (2) $\Omega_0=0.3$, $\mv=8.6\times10^{14}$ $M_{\odot}$, and 
  dark matter profile within \rv frozen for $z>z_{crit}$ (dashed line),
  and (3) $\Omega_0=0.3$, $\mv=1.3\times10^{15}$ $M_{\odot}$, and
  dark matter profile within \rv allowed to evolve as in the $\Omega_0=1$
  case (dotted line).\label{lz2}}
\end{center}
\end{figure}

\fig{lz2} compares the evolution of the luminosity for
two $\Omega_0=0.3$ models (which differ in the handling of the core
evolution as discussed above) and an $\Omega_0=1$ model, all with
$\mv\approx1.0\times10^{15}$ $M_{\odot}$ at $z=0$, and a
gas fraction of 0.3. Clearly the dependence of the luminosity evolution on
$\Omega_0$, and on the treatment of the core in the $\Omega_0=0.3$ models,
is slight. The weakness of the dependence on $\Omega_0$ comes about because
the luminosity is dominated by gas in the central regions, which has a 
very similar profile in the $\Omega_0=0.3$ 
and $\Omega_0=1$ cases. Significant differences occur at large radii,
but these regions contribute little to the X-ray flux.

\section{CONCLUSIONS\label{conc}}
The 1-D simulations presented above, including an evolving potential and
the effects of radiative cooling and distributed mass deposition, are
reasonably successful at reproducing many of the observed
properties of galaxy clusters. Temperature and density profiles
are broadly realistic, and the inflow of higher entropy gas into the
cluster core as a result of radiative cooling, leads to a gas fraction
which rises with radius, in accord with observations. It also 
results in X-ray luminosity evolution which is milder than the
self-similar prediction. The strong positive evolution of the latter is
known to conflict with observation (Castander \etal\ 1994).

There are, however, some significant discrepancies with observation.
The cooling flows produced by {\em EVOL} simulations are much stronger
than those typically observed. This result is robust to changes in the
values of numerical parameters in the simulations. The effects of
subcluster merging, which are believed to periodically disrupt cooling
flows, and are of course absent from a 1-D code, are likely to account for
the lower cooling rates (and the substantial scatter for
clusters of a given X-ray luminosity) seen in massive clusters.
However in lower mass systems, the cooling flow re-establishes itself 
very quickly after disruption.

A number of further disagreements with observation for low mass systems
suggest the need for additional physics in the models. The simulations
reported here do give somewhat flatter surface brightness profiles (low
$\beta$ values -- Figure~\ref{beta1}) and lower gas fractions
(Figure~\myref{frac1}) in low mass systems, but the effects are much
weaker than those observed. Also the slope of the $L:T$ relation does not
steepen below $T\sim$1~keV in the way which is observed
(\fig{lt1}). Perhaps most suggestive of all, is the observation
that the simple models described in this paper underpredict the entropy
of the gas in groups and poor clusters.

These differences suggest that some additional effects, which are most
significant in low mass systems, have increased the entropy and reduced the 
density of gas in the
inner regions of low mass systems. Possible mechanisms include heating of
the gas to a temperature $\gsim 10^7$~K, either as the result of an early
preheating era (Couchman \& Rees 1986; Meiksin \& Madau 1993) 
or the injection of galaxy
winds (White 1991; Metzler \& Evrard 1994), or increased efficiency of star
formation in groups (David, Jones \& Forman 1995).

The role of these additional mechanisms is explored in Paper II.

\section*{ACKNOWLEDGEMENTS}
We would like to thank Peter Thomas and Richard Bower for help and
discussions whilst this work was in progress, David White for
a prepublication copy of his compilation of cluster properties, and
Damian Cannon for the observed surface brightness profiles used in 
\sect{fgas}. Discussions with the referee, David Weinberg, 
resulted in improvements to the paper and to our understanding 
of self-similarity.
The computations were carried out using the facilities of the 
Starlink node at the University of Birmingham. PAK acknowledges the
support of the PPARC.

\parindent=0mm
\parskip=1mm
\let\doublespace=\relax
\def\sec@upcase#1{\relax{#1}}
\bgroup\parindent=\parindent\parskip=\parindent
\def\refpar{\par\hangindent=10mm\hangafter=1}
\def\reference{\relax\refpar}
\setlength{\baselineskip}{6mm}
\setlength{\parskip}{3mm}
\section*{REFERENCES}

\reference
Allen S.W., Fabian A.C., Johnstone R.M., Nulsen P.E.J., Edge A.C., 
  1992, \MNRAS{254}{51}

\reference
Allen S.W., Fabian A.C., Kneib J.P., 1996, \MNRAS{279}{615}

\reference
Bertschinger E., 1985, \ApJ{58}{39}

\reference
Carr B., 1994, \ARAA{32}{531}

\reference
Castander F.J., Ellis R.S., Frenk C.S., Dressler A., Gunn J.E.,
  1994, \ApJ{424}{L79}

\reference
Couchman H.M.P., Rees M.J., 1986, \MNRAS{221}{53}

\reference
Crone M.M., Evrard A.E., Richstone D.O., 1994, \ApJ{434}{402}

\reference
David L.P., Forman W., Jones C., 1990, \ApJ{359}{29}

\reference
David L.P., Jones C., Forman W., 1995, \ApJ{445}{578}

\reference
Efstathiou G., Eastwood J.W., 1981, \MNRAS{194}{503}

\reference
Evrard A.E., 1988, \MNRAS{235}{911}

\reference
Evrard A.E., 1990, \ApJ{363}{349}

\reference
Eyles C.J., Watt M.P., Bertram D., Church M.J., Ponman T.J.,
  Skinner G.K., Willmore A.P., 1991, \ApJ{376}{23}

\reference
Fabian A.C., Nulsen P.E.J., Canizares C.R., 1991, 
  {Astronomy and Astrophysics Review}, {2}, 191

\reference
Fillmore J.A., Goldreich P., 1984, \ApJ{281}{1} ~~(FG84)

\reference
Gingold R.A., Monaghan J.J., 1977, \MNRAS{181}{375}

\reference
Gull S.F., Northover K.J.E., 1975, \MNRAS{173}{585}

\reference
Gunn K.F., Thomas P.A., 1996, MNRAS, submitted 

\reference
Hirayama Y., 1978, {Prog. Theor. Phys.}, {60}, 724

\reference
Hoffman Y., Shaham J., 1985, \ApJ{297}{16}

\reference
Hughes J.P., Gorenstein P., Fabricant D., 1988, \ApJ{329}{82}

\reference
Jones C., Forman W., 1984, \ApJ{276}{38}
        
\reference
Kaiser N., 1986, \MNRAS{222}{323}

\reference
Katz N., White S.D.M., 1993, \ApJ{412}{455}

\reference
Kauffmann G., White S.D.M., 1993, \MNRAS{261}{921}

\reference
Knight P.A., 1996, Ph.D. thesis, University of Birmingham, UK

\reference
Lacey C., Cole S., 1993, \MNRAS{262}{627}

\reference
Lea S.M., 1976, \ApJ{203}{569}

\reference
Loewenstein M., Mushotzky R.F., 1996, \ApJSub

\reference
Markevitch M., 1996, \ApJ{465}{L1}

\reference
Markevitch M., Vikhlinin A., 1996, \ApJSub

\reference
Mathews W.G., Baker J.C., 1971, \ApJ{170}{241}

\reference
Meiksin A., Madau P., 1993, \ApJ{412}{34}

\reference
Metzler C.A., Evrard A.E., 1994, \ApJ{437}{564}

\reference
Mulchaey J.S., Davis D.S., Mushotzky R.F., Burnstein D., 1996,
  \ApJ{456}{80}

\reference
Mushotzky R., Loewenstein M., Arnaud K.A., Tamura T.,
  Fukazawa Y., Matsushita K., Kikuchi K., Hatsukade I.,
  1996, \ApJ{466}{686}

\reference
Navarro J.F., Frenk C.S., White S.D.M., 1996, \MNRAS{275}{720}

\reference
Navarro J.F., Frenk C.S., White S.D.M., 1996, \ApJ{462}{563} ~~(NFW96)

\reference
Oegerle W.R., Hoessel J.G., 1991, \ApJ{375}{15}

\reference
Ostriker J.P., Gnedin N.Y., 1996, \ApJSub

\reference
Perrenod S.C., 1978, \ApJ{224}{285}

\reference
Ponman T.J., Bertram D., 1993, \Nature{363}{51}

\reference
Ponman T.J., Bourner P.D.J., Ebeling H., B\"ohringer H., 1996, MNRAS,
in press

\reference
Raymond J.C, Cox D.P., Smith B.W., 1976, \ApJ{204}{290}

\reference
Richtmyer R.D., Morton K.W., 1967, {\em Difference Methods for Initial 
  Value Problems} (2d ed.; New York: Interscience)

\reference
Sarazin C.L., 1990, in {\em The Interstellar Medium in Galaxies}, ed
  H.A. Thronson Jr, J.M. Shull, Kluwer Academic

\reference
Schindler S., Hattori M., Neumann D.M., B\"ohringer H., 1996, A\&A,
submitted

\reference
Takahara F., Ikeuchi S., Shibazaki N., H\={o}shi R., 1976, {Prog.
  Theor. Phys.}, {56}, 1093

\reference
Thomas P.A., 1988, \MNRAS{235}{315}

\reference
Thomas P.A., Couchman H.M.P., 1992, \MNRAS{257}{11}

\reference
Thoul A.A., Weinberg D.H., 1995, \ApJ{442}{480}

\reference
Tormen G., Bouchet F.R., White S.D.M., 1996, MNRAS, submitted

\reference
Walker T.P., Steigman G., Kang H., Schramm D.M., Olive K.A., 
  1991, \ApJ{376}{51}

\reference
Watt M.P., Ponman T.J., Bertram D., Eyles C.J., Skinner G.K., 
  Willmore A.P., 1992, \MNRAS{258}{738}

\reference
Waxman E., Miralda-Escud\'{e} J., 1995, \ApJ{451}{451}

\reference
White D.A., Fabian A.C., 1995, \MNRAS{273}{72}

\reference
White D.A., Jones C., Forman W.R., 1996, MNRAS, submitted

\reference
White R.E., Sarazin C.L., 1987, \ApJ{318}{612}

\reference
White R.E., 1991, \ApJ{367}{69}

\reference
White S.D.M., 1976, \MNRAS{174}{19}

\reference
White S.D.M., Navarro J.N., Evrard A.E., Frenk C.S., 1993,
  \Nature{366}{429}

\reference
Yamashita K., 1992, in Tanaka Y., Koyama K. eds., {\em Frontiers of X-ray
Astronomy}. Universal Academy Press, Tokyo. p.475

\end{document}